\documentclass[preprint,prd,aps,showpacs,showkeys,nofootinbib]{revtex4}
\usepackage{graphicx}
\usepackage{amsmath}
\usepackage{hyperref}
\usepackage[T1]{fontenc}
\usepackage{soul}
\usepackage{color,xcolor}
\begin{document}

	
	\title{Lepton-flavor violation and two loop electroweak corrections to
		$(g-2)_\mu$ in the triplets
		next-to-minimal MSSM}
	
	\author{Zhao-Yang Zhang$^1$\footnote{1311274306@qq.com}}
	
	\affiliation{Department of Physics and Institute of Theoretical Physics,
		Nanjing Normal University, Nanjing,  210023, China$^1$\\}

	\begin{abstract}
		In the Standard Model (SM), charged lepton flavor-violating (LFV) processes are strictly forbidden, and thus observation of LFV would signal the presence of new physics. Recently, the Muon $g{-}2$ collaboration at Fermilab reported their final result for the muon magnetic dipole moment (MDM), which is now consistent with the latest SM prediction at the $1\sigma$ level. This imposes a significant constraint on new-physics contributions to $(g{-}2)_\mu$.
		In this work, we study the LFV processes $\ell_j^- \to \ell_i^-\gamma$ and $\ell_j^- \to \ell_i^- \ell_i^- \ell_i^+$ within the framework of the triplet next-to-minimal supersymmetric standard model (TNMSSM). We include two-loop contributions to the muon MDM and investigate the resulting constraints on the model parameter space. Our numerical analysis shows that, in the TNMSSM, doubly charged Higgs bosons and the couplings $Y_L$ and $Y_D$ associated with the Type-I+II seesaw mechanism play a crucial role in both LFV processes and the muon MDM. Moreover, two-loop electroweak corrections have a significant impact on the muon MDM in this framework.
		
	\end{abstract}

	\maketitle
	\section{Introduction\label{sec1}}
	\indent\indent
	Lepton flavor violation (LFV) has been extensively investigated in both theoretical and experimental studies, and remains a prominent topic in contemporary particle physics. The discovery of neutrino oscillations has demonstrated that lepton flavor is not fundamentally conserved~\cite{NeuExp,neu-b1}. However, in the Standard Model (SM), neutrinos are massless and lepton flavor is conserved~\cite{neu-b2,neu-b3,neu-b4}, which contradicts experimental observations and motivates extensions beyond the SM~\cite{neu-b5}. LFV thus provides a valuable probe for new physics~\cite{LFV7,LFV8,LFV9,LFV10,LFV11,HB2014,Zhao2015}. Table~\ref{tab1} summarizes the current experimental constraints and future sensitivities for the LFV decays $\ell_j^- \to \ell_i^- \gamma$ and $\ell_j^- \to \ell_i^- \ell_i^- \ell_i^+$~\cite{LFV1,LFV2,LFV3,LFV4,LFV5,LFV6,Huang2024}.
	
	Another important test of new physics is the anomalous magnetic dipole moment (MDM) of the muon, $(g{-}2)_\mu$~\cite{Peter2021,JL2018,JL2021}. The longstanding discrepancy between experimental results and SM predictions previously exceeded $4\sigma$. Recently, the Muon $g{-}2$ Collaboration at Fermilab reported their final measurement, $a_\mu = 1165920705(148) \times 10^{-12}$~\cite{Muong-2:2025xyk}, yielding a new world average $a_\mu^{\rm exp} = 1165920715(145) \times 10^{-12}$. Compared with the updated theoretical prediction $a_\mu^{\rm SM} = 116592033(62) \times 10^{-11}$~\cite{Muong-2s:2025xyk,Muong-2s1:2025xyk}, this results in 
	\begin{equation}
		\Delta a_\mu^{\rm new} = a_\mu^{\rm exp} - a_\mu^{\rm SM} = (39 \pm 64) \times 10^{-11},
		\label{g2n}\end{equation}
	which is within $1\sigma$, in contrast to the previous deviation $\Delta a_\mu^{\rm old} = (251 \pm 59) \times 10^{-11}$~\cite{Peter2021,Peter2,g23,g24}. While the new measurement tightens constraints on new physics, additional contributions to $(g-2)_\mu$ remain possible.
	
	In this work, we study charged LFV processes within the framework of the triplet next-to-minimal supersymmetric standard model (TNMSSM), considering both one-loop and two-loop contributions to the muon MDM. The TNMSSM extends the minimal supersymmetric standard model (MSSM) with two $SU(2)_L$ triplets and one singlet~\cite{TNM p1,ZJ1,ZJ2,YX}, addressing the little hierarchy problem and the $\mu$ problem simultaneously~\cite{up p1,hire p2,NMS p1,NMS p2,UH p1,UH p2,UH p3,UH p4,TM p1,TM p2,TM p3}.
	A triplet field with hypercharge $+1$, together with three right-handed neutrinos, generates tiny neutrino masses via a Type-I+II seesaw mechanism~\cite{ZJ1,ZJ2,type1 p1,type2 p1,type12 p1,type12-p2,type12-p3,type12-p4,type12-p5}. One-loop radiative corrections are included using the on-shell renormalization scheme to obtain effective neutrino masses consistent with observed oscillations~\cite{top-down,MASS DATA p1,UPMNS p1,UPMNS p2}. The triplet gives rise to doubly charged Higgs bosons, which dominantly mediate the tree-level $\ell_j^- \to \ell_i^- \ell_i^- \ell_i^+$ LFV processes, providing a distinctive signal of the TNMSSM.
	These processes can be probed in current and future experiments, including high-energy colliders and precision LFV searches~\cite{LSY1,LSY2,LSY3,LSY4}. In addition, one-loop and two-loop electroweak corrections, particularly those involving doubly charged Higgs bosons and their Yukawa couplings, significantly contribute to $(g{-}2)_\mu$. We consider both normal hierarchy (NH) and inverted hierarchy (IH) scenarios for neutrino masses.
	
	This paper is organized as follows: The framework of the TNMSSM, including the superpotential, soft breaking terms, and the mechanism generating tiny neutrino masses, is presented in Sec.~\ref{sec2}. In Sec.~\ref{sec3}, we present the decay widths of charged LFV processes and the muon  MDM. Sec.~\ref{sec4} contains the numerical analyses, and conclusions are given in Sec.~\ref{sec5}. Some key derivations and the mass matrices of relevant particles are collected in the Appendices for completeness.

	\begin{table*}
		\begin{tabular*}{\textwidth}{@{\extracolsep{\fill}}lll@{}}
			\hline
			LFV process & Present limit & Future sensitivity\\
			\hline
			$\mu\rightarrow e\gamma$ & $<4.2\times10^{-13}$ & $\sim6\times10^{-14}$ \cite{LFV1}\\
			$\mu\rightarrow 3e$ & $<1\times10^{-12}$ \cite{LFV2} & $\sim10^{-16}$ \cite{LFV3}\\
			$\tau\rightarrow e\gamma$ & $<3.3\times10^{-8}$ \cite{LFV4} & $\sim10^{-8}-10^{-9}$ \cite{LFV5}\\
			$\tau\rightarrow 3e$ & $<2.7\times10^{-8}$ \cite{LFV6} & $\sim10^{-9}-10^{-10}$\\
			$\tau\rightarrow \mu\gamma$ & $<4.2\times10^{-8}$ & $\sim10^{-8}-10^{-9}$\\
			$\tau\rightarrow 3\mu$ & $<2.1\times10^{-8}$ & $\sim10^{-9}-10^{-10}$\\
			\hline
		\end{tabular*}
		\caption{Present constraints and future sensitivities on the branching ratios of LFV processes. }
		\label{tab1}
	\end{table*}
	\section{The TNMSSM}
	\label{sec2}
	Besides the MSSM superfields, the TNMSSM includes a gauge singlet $S$ and two $SU(2)_L$ triplets $T$ and $\bar T$.  
	To address the $\mu$-problem, a discrete $Z_3$ symmetry is imposed, under which all superfields 
	($\hat H_u,\hat H_d,\hat S,\hat T,\hat{\bar T}$) carry charge $1 \pmod 3$~\cite{TNM p1,ZJ1,ZJ2}.  
	This forbids bilinear terms such as $\mu\,\hat H_u \cdot H_d$ and allows only cubic terms like  
	$\lambda\,\hat S \hat H_u\!\cdot\!\hat H_d$ and  
	$\lambda_T\,\hat S\,\text{Tr}(\hat T \hat{\bar T})$.  
	After $\hat S$ acquires a vacuum expectation value, an effective $\mu$-term $\mu_{\rm eff} = \lambda \langle S \rangle$ is dynamically generated.
	The corresponding superpotential reads~\cite{ZJ1,ZJ2,YX}:
	\begin{align}
		W_{\rm TNMSSM} &= S \left( \lambda H_u\cdot H_d + \lambda_T \textnormal{tr}(\bar{T} T) \right)
		+ \frac{\kappa}{3} S^3
		+ \chi_u H_u \cdot \bar{T} H_u
		+ \chi_d H_d \cdot T H_d \nonumber\\
		&\quad + h_u H_u\cdot Q\bar{u}+h_d H_d\cdot Q\bar{d}+h_eH_d\cdot L \bar{e}.
		\label{eq:tnmssm_superpotential}
	\end{align}
	
	Adding three singlet right-handed neutrino superfields $N^c$, the superpotential extends to ~\cite{ZJ1}
	\begin{equation}
		W_{\rm TNMSSM+Seesaw} = W_{\rm TNMSSM} 
		+ Y_L\, L\!\cdot\! T L
		+ Y_D\, L\!\cdot\! H_u N^c
		+ Y_R\, N^c S N^c,
		\label{eq:type1+2_superpotential}
	\end{equation}
	where $Y_{L,D,R}$ are the corresponding Yukawa matrices.
	
	The soft SUSY-breaking Lagrangian is
	\begin{eqnarray}
		-\mathcal{L}_{\text{soft}}
		&=& m_{H_u}^2|H_u|^2 + m_{H_d}^2|H_d|^2 + m_{S}^2|S|^2
		+ m_{T}^2 \mathrm{tr}|T|^2 + m_{\bar T}^2 \mathrm{tr}|\bar T|^2
		+ m_{N^c}^2|N^c|^2 \nonumber \\
		&& + m_Q^2|Q|^2 + m_{\bar u}^2|\bar u|^2 + m_{\bar d}^2|\bar d|^2
		+ m_{\tilde L}^2|L|^2 + m_{\tilde e}^2|\tilde e|^2 \nonumber \\
		&& + \bigl(
		A_{h_u} Q\!\cdot\! H_u \bar u
		- A_{h_d} Q\!\cdot\! H_d \bar d
		- A_{h_e} L\!\cdot\! H_d \bar e
		+ A\, S H_u\!\cdot\! H_d
		+ A_T S \mathrm{tr}(T\bar T)
		+ \tfrac{A_\kappa}{3} S^3 \nonumber \\
		&& \qquad
		+ A_u H_u\!\cdot\!\bar T H_u
		+ A_d H_d\!\cdot\! T H_d
		+ A_{Y_L} L\!\cdot\! T L
		+ A_{Y_N} L\!\cdot\! H_u N^c
		+ A_{Y_R} N^c S N^c
		+ \text{h.c.}\bigr) \nonumber \\
		&& - \tfrac{1}{2}\Bigl(
		M_3 \tilde\lambda_3 \tilde\lambda_3
		+ M_2 \tilde\lambda_2 \tilde\lambda_2
		+ M_1 \tilde\lambda_1 \tilde\lambda_1
		+ \text{h.c.}\Bigr).
		\label{eq:new}
	\end{eqnarray}
	
	The neutral scalar fields in the Higgs and triplet sectors can be expanded around their vacuum expectation values (VEVs) as
	\begin{eqnarray}
		H_d^0 &=& \frac{1}{\sqrt{2}} \bigl( v_d + h_d + i P_d \bigr), \qquad
		H_u^0 = \frac{1}{\sqrt{2}} \bigl( v_u + h_u + i P_u \bigr), \nonumber\\
		T^0 &=& \frac{1}{\sqrt{2}} \bigl( v_T + h_T + i P_T \bigr), \qquad
		\bar{T}^0 = \frac{1}{\sqrt{2}} \bigl( v_{\bar{T}} + h_{\bar{T}} + i P_{\bar{T}} \bigr), \nonumber\\
		S^0 &=& \frac{1}{\sqrt{2}} \bigl( v_S + h_S + i P_S \bigr),
	\end{eqnarray}
	where $h_i$ and $P_i$ denote the CP-even and CP-odd components of each field, respectively, and $v_i$ are their corresponding VEVs.  
	
	It is convenient to define the combined VEVs, ratios, and the effective $\mu$ parameters as
	\begin{gather}
		v_{ud}^2 = v_u^2 + v_d^2, \quad 
		v_{T\bar T}^2 = v_T^2 + v_{\bar T}^2, \\
		\tan\beta = \frac{v_u}{v_d}, \quad
		\tan\beta' = \frac{v_T}{v_{\bar T}}, 
	\end{gather}
	\begin{equation}
		\mu = \lambda v_S, \qquad 
		\mu_T = \lambda_T v_S.
		\label{uuuu}
	\end{equation}

	Products of $SU(2)_L$ multiplets are understood as the antisymmetric contraction of their doublet indices, and can be written explicitly as~\cite{TNM p1}:
	\begin{align}
		H_u \!\cdot\! H_d
		&= H_u^+ H_d^- - H_u^0 H_d^0, \label{eq:hu_hd} \\
		H_u \!\cdot\! \bar T H_u
		&= \sqrt{2}\, H_u^+ H_u^0 \bar T^- - (H_u^0)^2 \bar T^0 - (H_u^+)^2 \bar T^{--}, \label{eq:hu_tbar_hu} \\
		H_d \!\cdot\! T H_d
		&= \sqrt{2}\, H_d^- H_d^0 T^+ - (H_d^0)^2 T^0 - (H_d^-)^2 T^{++}, \label{eq:hd_t_hd} \\
		Y_L\, L \!\cdot\! T L
		&= Y_L^{ij}\!\bigl(\nu_i\nu_j T^0
		+ \tfrac{1}{\sqrt{2}}(\nu_i\ell_j+\ell_i\nu_j)T^+
		- \ell_i\ell_j T^{++}\bigr), \label{eq:lltl} \\
		Y_D\, L \!\cdot\! H_u N^c
		&= Y_D^{ij}\!\bigl(\nu_i H_u^0 - \ell_i H_u^+\bigr) N_j^c. \label{eq:yd_hu_n}
	\end{align}

	The complete $6\times6$ neutrino mass matrix in the TNMSSM, which simultaneously realizes a combined Type-I and Type-II
		seesaw mechanism, is~\cite{onel1,onel2,yao,Zhao1,Yan1}
	\begin{equation}
		\label{eq:Mnu}
		M_\nu =
		\begin{pmatrix}
			M_L & M_D \\
			M_D^{T} & M_R
		\end{pmatrix},
	\end{equation}
	where $M_D$ is the $3\times3$ Dirac mass matrix, and $M_L$ and $M_R$ are the Majorana mass matrices for the left- and right-handed neutrinos, respectively.  
	They are generated at tree level by
	\begin{equation}
		\label{MLR}
		M_L = \sqrt{2}\,Y_L v_T,\qquad
		M_D = \frac{Y_D v_u}{\sqrt{2}},\qquad
		M_R = \sqrt{2}\,Y_R v_S .
	\end{equation}
	
	Diagonalization of the full matrix is achieved by a unitary matrix $Z_{N_\nu}$,
	\begin{equation}
		Z_{N_\nu}^T M_\nu Z_{N_\nu}
		= \mathrm{diag}(m_{\nu_1},\dots , m_{\nu_6}) ,
	\end{equation}
	which can be approximated at leading order in
	$\varsigma \equiv M_D M_R^{-1}$ as~\cite{yao,Zhao1}
	\begin{equation}
		\label{SR}
		Z_{N_\nu}^T =
		\begin{pmatrix} \mathcal{S}^T & 0 \\ 0 & \mathcal{R}^T \end{pmatrix}
		\begin{pmatrix}
			1 - \tfrac12 \varsigma^\dag \varsigma & -\varsigma^\dag \\
			\varsigma & 1 - \tfrac12 \varsigma \varsigma^\dag
		\end{pmatrix}.
	\end{equation}
	Here $\mathcal{S}$ and $\mathcal{R}$ diagonalize the light and heavy
	sub-blocks,
	\begin{equation}
		\mathcal{S}^T M_\nu^{\text{tree}} \mathcal{S}
		= \mathrm{diag}(m_{\nu_1}, m_{\nu_2}, m_{\nu_3}),\qquad
		\mathcal{R}^T M_R \mathcal{R}
		= \mathrm{diag}(m_{\nu_4}, m_{\nu_5}, m_{\nu_6}) .
	\end{equation}
	
	At tree level the effective light-neutrino mass matrix obtained
	from Eq.~(\ref{eq:Mnu}) is
	\begin{equation}
		\label{meff1}
		M_\nu^{\text{tree}}
		\simeq m_\nu^{II}+m_\nu^{I}
		= M_L - M_D M_R^{-1} M_D^{T},
	\end{equation}
	showing the combined Type-II and Type-I seesaw contributions.

	We then include one-loop radiative corrections, using the on-shell scheme to cancel ultraviolet divergences~\cite{onel2,yao}.  
	The ${\nu_i}^0$–${\nu_j}^0$ mixing arises from loops of three types: gauge bosons, Higgs bosons (neutral and charged), and supersymmetric scalars (sneutrinos or sleptons) coupled with neutralinos or charginos.  
	The correction can be expressed as
	\begin{eqnarray}
		\delta m_{\nu,ij}^{1\text{-loop}}
		&=& \delta m_{\nu,ij}^{(Z,\nu^0)}
		+ \delta m_{\nu,ij}^{(W,e^-)}
		+ \delta m_{\nu,ij}^{(H^0,\nu^0)}
		+ \delta m_{\nu,ij}^{(A^0,\nu^0)} \nonumber \\
		&& + \delta m_{\nu,ij}^{(\tilde{\nu}^I,\tilde{\chi})}
		+ \delta m_{\nu,ij}^{(\tilde{\nu}^R,\tilde{\chi})}
		+ \delta m_{\nu,ij}^{(H^+,e^-)}
		+ \delta m_{\nu,ij}^{(\tilde{e},\chi^-)} .
		\label{ONE}
	\end{eqnarray}
	Here, $\delta m_{\nu,ij}^{1\text{-loop}}$ denotes the one-loop correction, with details given in Appendix~\ref{oloop}.

	With the one-loop corrections, the neutrino mass matrix defined in
	Eq.~(\ref{eq:Mnu}) becomes
	\begin{eqnarray}
		M_{\nu}^{\rm sum}
		&=& M_{\nu} + Z_{N_\nu} \, \delta m_{\nu} \, Z_{N_\nu}^T  \nonumber\\
		&=&
		\begin{pmatrix}
			(M_L+\delta m_L)_{3\times3} & (M_D+\delta m_D)_{3\times3} \\
			(M_D+\delta m_D)^T_{3\times3} & (M_R+\delta m_R)_{3\times3}
		\end{pmatrix},
		\label{NUMOL}
	\end{eqnarray}
	where \(M_\nu\) is given in Eq.~(\ref{eq:Mnu}).
	The matrix $M_{\nu}^{\rm sum}$ retains the seesaw structure. 
	Analogous to Eq.~(\ref{meff1}), the one-loop corrected effective light neutrino mass matrix is given by~\cite{ZJ1,ZJ2}  
	\begin{eqnarray}  
		M_{\nu}^{\rm eff} \approx (M_L + \delta m_L) - (M_D + \delta m_D) (M_R + \delta m_R)^{-1} (M_D + \delta m_D)^T.
		\label{meff}  
	\end{eqnarray}  
	Using the "top-down" method~\cite{top-down,neu-b5,Yan1}, $M_{\nu}^{\rm eff}$ can be diagonalized to obtain the three light neutrino masses and PMNS mixing angles.

	In the TNMSSM, tiny neutrino masses arise from both Type-I and Type-II seesaw mechanisms. An interesting case occurs when the Type-II contribution, including its one-loop correction, 
	$M_L + \delta m_L$, is comparable in magnitude to the Type-I term 
	$(M_D + \delta m_D)(M_R + \delta m_R)^{-1}(M_D + \delta m_D)^{T}$~\cite{CW1,CW2,CW3}.
	This partial cancellation naturally explains the small neutrino masses without requiring extremely suppressed Yukawa couplings $Y_L$ and $Y_D$, enhancing their phenomenological impact, particularly in LFV processes. In our numerical analysis (Section~\ref{sec4}), the full neutrino mass matrix is obtained by combining $M_\nu^{\rm tree}$ with $\delta m_\nu^{1\text{-loop}}$, ensuring consistency with experimental constraints on neutrino masses and PMNS mixing angles.

	\section{Lepton flavor violation and $(g-2)_\mu$ in the TNMSSM}
	\label{sec3}
	In this section, we compute and analyze the LFV processes
	$\ell_{j}^{-}\to \ell_{i}^{-}\gamma$ and $\ell_{j}^{-}\to \ell_{i}^{-}\ell_{i}^{-}\ell_{i}^{+}$. 
	Both channels are calculated at the one-loop level, while the three-body decay
	$\ell_{j}^{-}\to \ell_{i}^{-}\ell_{i}^{-}\ell_{i}^{+}$ also receives tree-level contributions. 
	In addition, we discuss the impact of potential new-physics two-loop diagrams on the 
	muon  MDM.
	
	\subsection{Rare decay $\ell_j^- \to \ell_i^- \gamma$}
	
\begin{figure}
		\setlength{\unitlength}{1mm}
		\centering
		\includegraphics[width=4.5in]{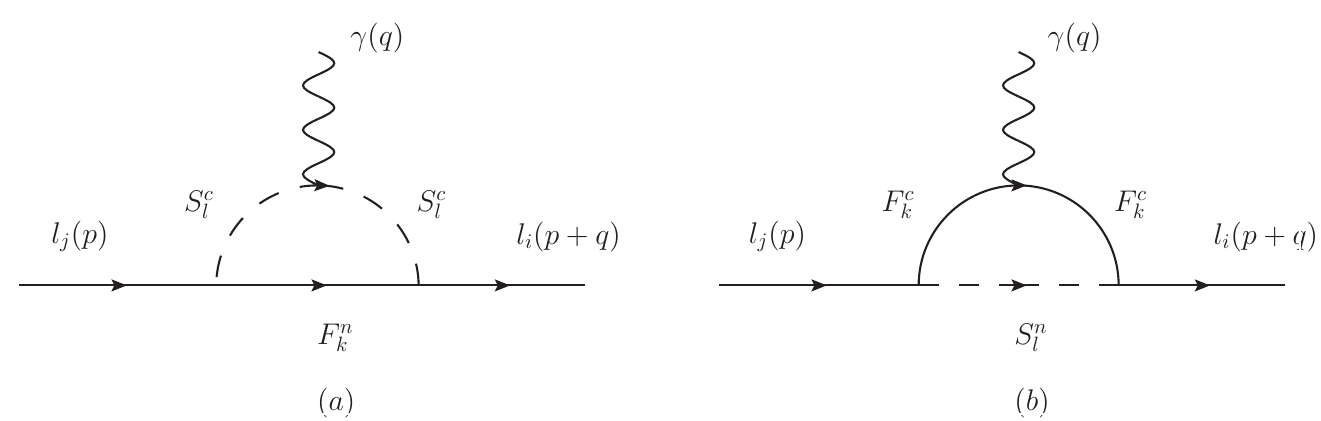}
		\vspace{0cm}
		\caption[]{One-loop diagrams contributing to the process $\ell_j^- \to \ell_i^- \gamma$ in the TNMSSM. In panel (a), $F^n_k$ and $S^c_l$ represent the neutral fermion and charged scalar, respectively. In panel (b), $F^c_k$ and $S^l_l$ denote the charged fermion and neutral scalar, respectively.
		}
		\label{FeynmanA}
	\end{figure}

	The off-shell amplitude for the process $\ell_j^- \to \ell_i^- \gamma$ can be expressed in the general form~\cite{JL2018,LLV1}
	\begin{align}
		T
		&= e\,\epsilon^{\mu}\,\bar{u}_{i}(p+q)
		\left[
		q^{2}\gamma_{\mu}\left(A_{1}^{L}P_{L}+A_{1}^{R}P_{R}\right)
		+ m_{\ell_j} i\sigma_{\mu\nu} q^{\nu}\left(A_{2}^{L}P_{L}+A_{2}^{R}P_{R}\right)
		\right]
		u_{j}(p),
		\label{llr}
	\end{align}
	where $p$ and $q$ denote the four-momenta of the incoming lepton $\ell_j^-$ 
	and the photon $\gamma$, respectively.  
	Here $\epsilon^{\mu}$ is the photon polarization vector, $P_{L}$ and $P_{R}$ are the standard chiral projection operators, and $u_i$ is the spinor wave function of the lepton.
	The one-loop Feynman diagrams contributing to the LFV decay $\ell_j^- \to \ell_i^- \gamma$ are shown in Fig.~\ref{FeynmanA}. The evaluation of these diagrams yields the coefficients $A_{1,2}^{L,R}$ appearing in Eq.~(\ref{llr}):
	\begin{eqnarray}
		&&A_1^{L,R}=A_1^{(a)L,R}+A_1^{(b)L,R},\nonumber\\
		&&A_2^{L,R}=A_2^{(a)L,R}+A_2^{(b)L,R},
	\end{eqnarray}
	where the explicit expressions for $A_{1,2}^{(a)L,R}$ and $A_{1,2}^{(b)L,R}$ are listed in Appendix~\ref{wilsonllr}.
	Combining the amplitude in Eq.~(\ref{llr}), the decay width for the process $\ell_j^- \to \ell_i^- \gamma$ can be expressed as
	\begin{align}
		\Gamma \left(\ell_j^- \to \ell_i^- \gamma \right) = \frac{e^2}{16\pi} , m_{\ell_j}^5 \left( |A_2^L|^2 + |A_2^R|^2 \right).
	\end{align}
	The corresponding branching ratio is then given by
	\begin{align}
		\text{Br}\left(\ell_j^- \to \ell_i^- \gamma \right) = \frac{\Gamma \left( \ell_j^- \to \ell_i^- \gamma \right)}{\Gamma_{\ell_j^-}}.
	\end{align}
	Here, $\Gamma_{\ell_j^-}$ denotes the total decay width of the lepton $\ell_j^-$. For numerical evaluations, we approximately use $\Gamma_\mu \simeq 2.996 \times 10^{-19}~\mathrm{GeV}$ for the muon and $\Gamma_\tau \simeq 2.265 \times 10^{-12}~\mathrm{GeV}$ for the tauon~\cite{PDG}.

	\subsection{Rare decay $\ell_j^- \to \ell_i^- \ell_i^- \ell_i^+$}
\begin{figure}
		\setlength{\unitlength}{1mm}
		\centering
		\includegraphics[width=2.5in]{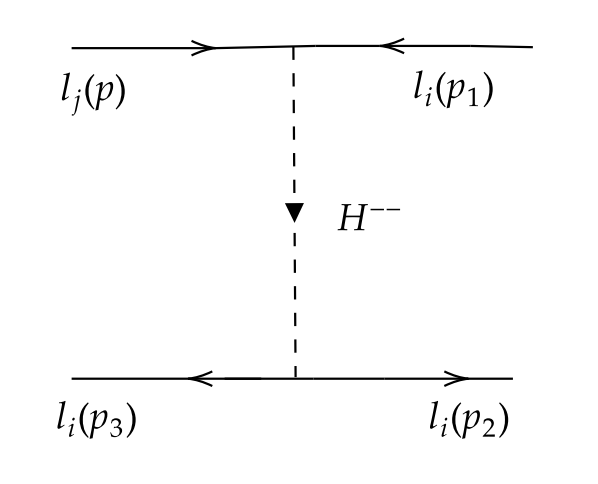}
		\vspace{0cm}
		\caption[]{The tree-level Feynman diagrams for the process $\ell_j^- \to \ell_i^- \ell_i^- \ell_i^+$, mediated by the doubly-charged Higgs bosons $H^{--}$ in the TNMSSM.}
		\label{FeynmanT}
	\end{figure}

	The effective amplitude for the process $\ell_j^- \to \ell_i^- \ell_i^- \ell_i^+$ receives contributions from both tree-level and one-loop diagrams.
	At tree level, this process is mediated by a doubly charged Higgs boson, as illustrated in Fig.~\ref{FeynmanT}. The corresponding amplitude reads~\cite{Huang2024}
	\begin{align}
		T &= C_{l_i l_j H^{c--*}}^L \;
		C_{\bar l_i \bar l_i H^{c--}}^R \;
		\frac{i}{(p - p_1)^2 - m_{H^{--}}^2} \;
		\bar u_i(p_1) P_L u_j(p) \;
		\bar u_i(p_3) P_R v_i(p_2) .
	\end{align}
	Here, $m_{H^{--}}$ denotes the mass of the doubly charged Higgs boson, while $C_{l_i l_j H^{c--*}}^L$ and $C_{\bar l_i \bar l_i H^{c--}}^R$ are the corresponding left- and right-handed couplings to the leptons.
	\begin{figure}
		\setlength{\unitlength}{1mm}
		\centering
		\includegraphics[width=2.5in]{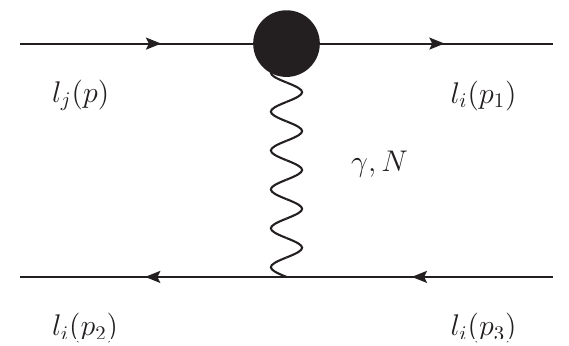}
		\vspace{0cm}
		\caption[]{The penguin-type diagrams of the $\ell_j^- \to \ell_i^- \ell_i^- \ell_i^+$ process, where $N$ denotes the $Z$ boson, and the large black dot indicates either the $\ell_j^- \ell_i^- \gamma$ or the $\ell_j^- \ell_i^- Z$ effective vertex.}
		\label{figl3lpg}
	\end{figure}
	
	We now consider the one-loop contributions.
	At this order, the effective amplitude of the LFV processes generally receives corrections from both penguin-type and box-type diagrams.
	As illustrated in Fig.~\ref{figl3lpg}, the penguin diagrams can be classified into two types: those mediated by a photon (photon-type) and those mediated by a massive vector boson (massive-vector type).
	The large black dot in Fig.~\ref{figl3lpg} represents a complete set of one-loop triangle diagrams, which are of the same type as those shown in Fig.~\ref{FeynmanA}. In contrast to Fig.~\ref{FeynmanA}, which contains only photon–lepton vertices, the large black dot in Fig.~\ref{figl3lpg} also includes the couplings to the $Z$ boson.
	This compact notation preserves the physical content of both photon and $Z$ boson interactions while keeping the diagram visually concise, without omitting the underlying loop structure.
	
	The invariant amplitude for the $\gamma$–penguin contribution is given by~\cite{JL2018,Huang2024,LLV1}:
	\begin{eqnarray}
		&&T_{\gamma-{\rm penguin}}=\bar u_i(p_1)[q^2 \gamma_\mu(A_1^LP_L+A_1^RP_R)+m_{l_j}i\sigma_{\mu\nu}q^\nu(A_2^LP_L+A_2^RP_R)]u_j(p)\nonumber\\
		&&\qquad\qquad\quad\times\frac{e^2}{q^2}\bar u_i(p_2)\gamma^\mu \nu_i(p_3)-(p_1\leftrightarrow p_2).\label{rpenguin}
	\end{eqnarray}
	Similarly, the invariant amplitude for the $N$-type penguin is written as:
	\begin{eqnarray}
		&&T_{N-{\rm penguin}}=\frac{e^2}{m_N^2}\bar u_i(p_1)\gamma_\mu(F^LP_L+F^RP_R)u_j(p)\bar u_i(p_2)\gamma^\mu(C_{\bar l_iNl_i}^LP_L\nonumber\\
		&&\qquad\qquad\quad+C_{\bar l_iNl_i}^RP_R)\nu_i(p_3)-(p_1\leftrightarrow p_2).\label{Npenguin}
	\end{eqnarray}	
	Here, $m_N$ denotes the mass of the $Z$ boson, and the explicit expressions for $F^{L,R}$ are given in Appendix~\ref{wilsonllr}.

\begin{figure}
		\setlength{\unitlength}{1mm}
		\centering
		\includegraphics[width=4.5in]{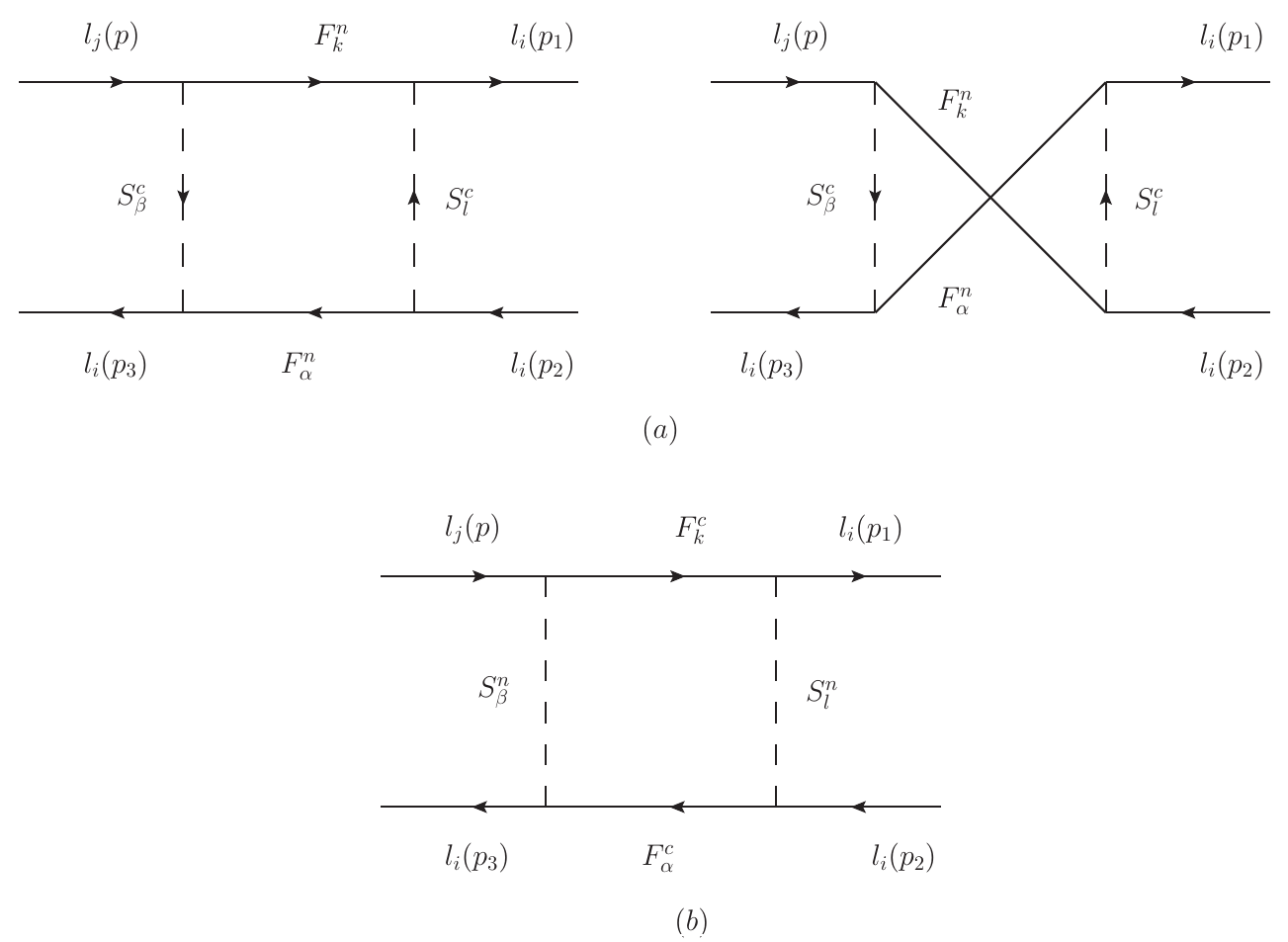}
		\vspace{0cm}
		\caption[]{Box-type diagrams for the process $\ell_j^- \to \ell_i^- \ell_i^- \ell_i^+$. 
			In diagram (a), $F^n_k$ and $S^c_l$ denote the neutral fermion and charged scalar, 
			respectively. In diagram (b), $F^c_k$ and $S^l_l$ denote the charged fermion and 
			neutral scalar, respectively.
		}
		\label{figl3lBX}
	\end{figure}

	The box-type diagrams also provide significant contributions to the 
	$\ell_j^- \to \ell_i^- \ell_i^- \ell_i^+$ process.  
	The corresponding Feynman diagrams are shown in Fig.~\ref{figl3lBX}.  
	The effective amplitude can be written as~\cite{JL2018,Huang2024}
	\begin{eqnarray}
		T_{\mathrm{box}} &=&
		\Big\{ B_1^L e^2 \,\bar u_i(p_1)\gamma_\mu P_L u_j(p)\,
		\bar u_i(p_2)\gamma^\mu P_L \nu_i(p_3)
		+ (L \leftrightarrow R) \Big\} \nonumber\\
		&& + \Big\{ B_2^L \big[ e^2 \,\bar u_i(p_1)\gamma_\mu P_L u_j(p)\,
		\bar u_i(p_2)\gamma^\mu P_R \nu_i(p_3)
		- (p_1 \leftrightarrow p_2) \big]
		+ (L \leftrightarrow R) \Big\} \nonumber\\
		&& + \Big\{ B_3^L \big[ e^2 \,\bar u_i(p_1) P_L u_j(p)\,
		\bar u_i(p_2) P_R \nu_i(p_3)
		- (p_1 \leftrightarrow p_2) \big]
		+ (L \leftrightarrow R) \Big\} \nonumber\\
		&& + \Big\{ B_4^L \big[ e^2 \,\bar u_i(p_1) \sigma_{\mu\nu} P_L u_j(p)\,
		\bar u_i(p_2) \sigma^{\mu\nu} P_L \nu_i(p_3)
		- (p_1 \leftrightarrow p_2) \big]
		+ (L \leftrightarrow R) \Big\},
	\end{eqnarray}
	where the coefficients $B_{1,2,3,4}^{L,R}$ are obtained from the 
	box-diagram calculations (see Appendix~\ref{wilsonllr}).  
	The decay width at one-loop order then follows as~\cite{JL2018,Huang2024}:
	
	\begin{eqnarray}
		&&\Gamma(l_j^-\rightarrow l_i^-l_i^-l_i^+)=\frac{e^4m_{l_j}^5}{512\pi^3}\Big\{(|A_2^L|^2+|A_2^R|^2)\Big(\frac{16}{3}\ln\frac{m_{l_j}}{2m_{l_i}}-
		\frac{14}{9}\Big)\nonumber\\
		&&\qquad\quad\qquad\quad\qquad+(|A_1^L|^2+|A_1^R|^2)-2(A_1^LA_2^{R*}+A_2^LA_1^{R*}+H.c.)+\frac{1}{6}(|B_1^L|^2+|B_1^R|^2)
		\nonumber\\
		&&\qquad\quad\qquad\quad\qquad+\frac{1}{3}(|B_2^L|^2+|B_2^R|^2)+\frac{1}{24}(|B_3^L|^2+|B_3^R|^2)+6(|B_4^L|^2+|B_4^R|^2)
		\nonumber\\
		&&\qquad\quad\qquad\quad\qquad-\frac{1}{2}(B_3^LB_4^{L*}+B_3^RB_4^{R*}+H.c.)+\frac{1}{3}(A_1^LB_1^{L*}+A_1^RB_1^{R*}+
		A_1^LB_2^{L*}\nonumber\\
		&&\qquad\quad\qquad\quad\qquad+A_1^RB_2^{R*}+H.c.)-\frac{2}{3}(A_2^RB_1^{L*}+A_2^LB_1^{R*}+A_2^LB_2^{R*}+A_2^RB_2^{R*}+
		H.c.)\nonumber\\
		&&\qquad\quad\qquad\quad\qquad+\frac{1}{3}\Big[2(|F^{LL}|^2+|F^{RR}|^2)+(|F^{LR}|^2+|F^{RL}|^2)+(B_1^LF^{LL*}+
		B_1^RF^{RR*}\nonumber\\
		&&\qquad\quad\qquad\quad\qquad+B_2^LF^{LR*}+B_2^RF^{RL*}+H.c.)+2(A_1^LF^{LL*}+A_1^RLF^{RR*}+H.c.)\nonumber\\
		&&\qquad\quad\qquad\quad\qquad+(A_1^LF^{LR*}+A_1^RLF^{RL*}+H.c.)-4(A_2^RF^{LL*}+A_2^LF^{RR*}+H.c.)\nonumber\\
		&&\qquad\quad\qquad\quad\qquad-2(A_2^LF^{RL*}+A_2^RLF^{LR*}+H.c.)\Big]\Big\},
	\end{eqnarray}
	where
	\begin{align}
		& {{F}^{LL}}=\underset{N=Z}{\mathop \sum }\,\frac{{{F}^{L}}C_{\overline{{{l}_{i}}}N{{l}_{i}}}^{L}}{m_{N}^{2}},\text{ }{{F}^{RR}}={{F}^{LL}}\left( L\leftrightarrow R \right),\text{ }\nonumber \\ 
		& {{F}^{LR}}=\underset{N=Z}{\mathop \sum }\,\frac{{{F}^{L}}C_{\overline{{{l}_{i}}}N{{l}_{i}}}^{R}}{m_{N}^{2}},\text{ }{{F}^{LR}}={{F}^{RL}}\left( L\leftrightarrow R \right).  
	\end{align}
	\subsection{$(g-2)_\mu$}
	
	Finally, we further analyze the muon  MDM. The latest discrepancy between the Standard Model (SM) theoretical prediction and the experimental measurement, as shown in Eq.~\ref{g2n}, is a $1\sigma$ deviation. This updated result suggests that the SM may now reasonably account for the long-standing muon $g-2$ anomaly. However, this does not definitively rule out the existence of potential new physics beyond the SM, a topic that remains a compelling target for both theoretical and experimental investigation.
	In general, the MDM of a Dirac fermion can be expressed through the effective Lagrangian as~\cite{EFT1,Feng,Feng1,Feng2,Zhang1}
	\begin{eqnarray}
		&&\mathcal{L}_{MDM}=\frac{e}{4m_\mu}a_\mu\bar l_\mu\sigma^{\alpha\beta}l_\mu F_{\alpha\beta},
		\label{LMDM}
	\end{eqnarray}
	where $\sigma^{\alpha\beta}=\tfrac{i}{2}[\gamma^\alpha,\gamma^\beta]$, $F_{\alpha\beta}$ denotes the electromagnetic field strength, $m_\mu$ is the muon mass, $a_\mu=\tfrac{1}{2}(g-2)_\mu$, and $l_\mu$ is the muon field. Using the effective Lagrangian method, we obtain the expression for $a_\mu$ as~\cite{Feng,Feng1,Feng2,Zhang1}
	\begin{eqnarray}
		&&a_\mu=\frac{4Q_fm_\mu^2}{(4\pi)^2}\Re(C_2^R+C_2^{L*}+C_6^R),
		\label{MDM}
	\end{eqnarray}
	where $Q_f=-1$, and $C_{2,6}^{L,R}$ are the Wilson coefficients of the operators $O_{2,6}^{L,R}$~\cite{ZJ1,Zhang1}.
	\begin{eqnarray}
		O_2^{L,R} &=& \frac{e Q_f}{(4\pi)^2}(-i D_\alpha^*)\, 
		\bar l_\mu \gamma^\alpha (F\!\cdot\!\sigma) P_{L,R} l_\mu , \nonumber\\
		O_6^{L,R} &=& \frac{e Q_f m_\mu}{(4\pi)^2}\,
		\bar l_\mu (F\!\cdot\!\sigma) P_{L,R} l_\mu ,
	\end{eqnarray}
	where $\mathcal{D}_\mu=\partial_\mu + i e A_\mu$, 
	$P_L=\tfrac{1}{2}(1-\gamma_5)$, and $P_R=\tfrac{1}{2}(1+\gamma_5)$.  
	From the diagrams in Fig.~\ref{FeynmanA}, the one-loop contribution to the muon MDM is
	\begin{eqnarray}
		a_\mu^{\text{one-loop}} = a_\mu^{(a)} + a_\mu^{(b)},
		\label{oneloop MDM}
	\end{eqnarray}
	where $a_\mu^{(a)}$ and $a_\mu^{(b)}$ correspond to the contributions from 
	Figs.~\ref{FeynmanA}(a) and \ref{FeynmanA}(b), respectively.  
	Their explicit expressions are given in Appendix~\ref{au}.
	
	\begin{figure}
		\setlength{\unitlength}{1mm}
		\centering
		\includegraphics[width=6.5in]{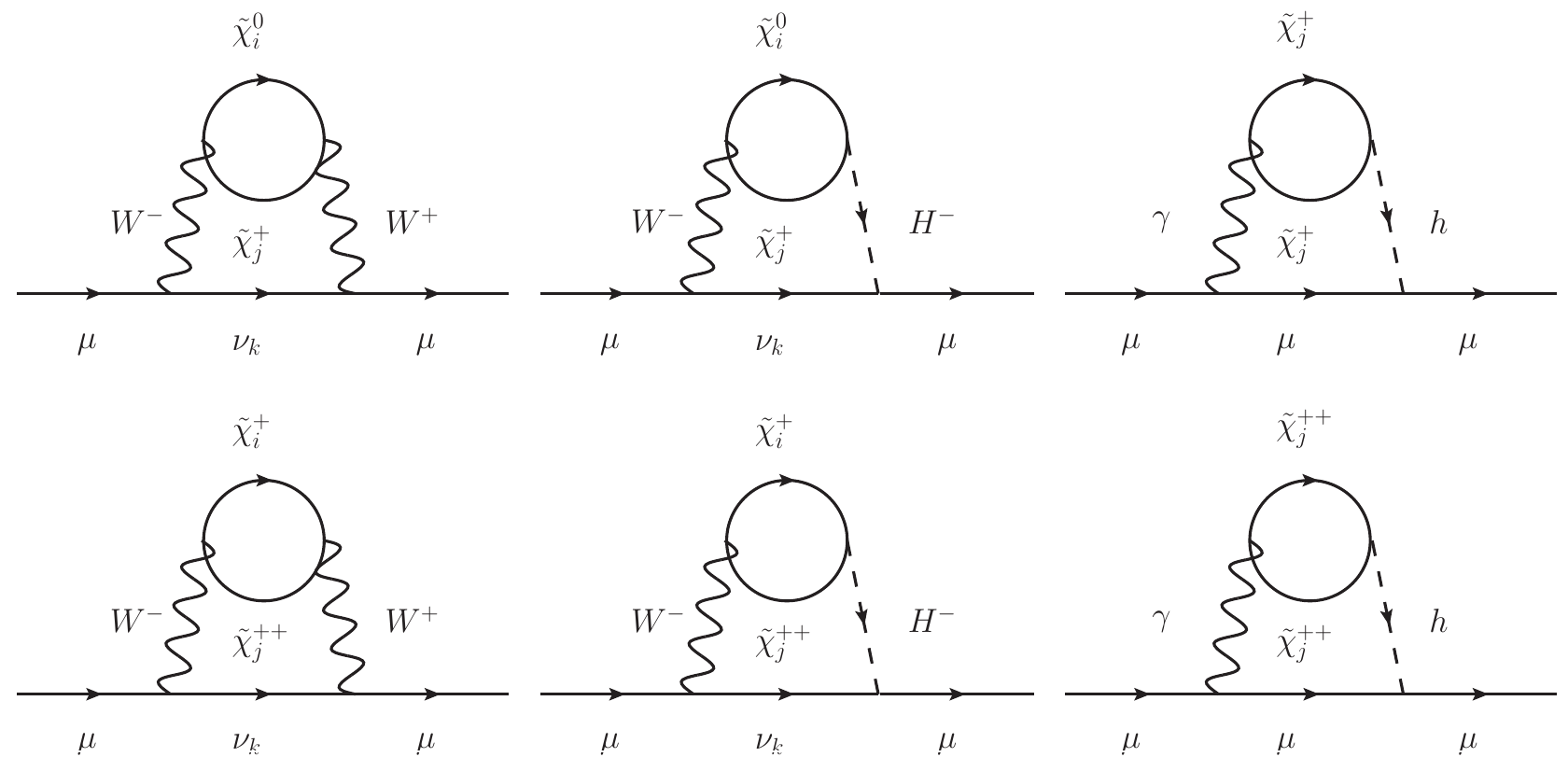}
		\vspace{0cm}
		\caption[]{Two-loop Barr--Zee type diagrams contributing to $a_\mu$ via photon insertions on internal lines.
			
		}
		\label{BZT}
	\end{figure}
	In a reasonable parameter space, the two-loop Barr–Zee diagrams shown in Fig.~\ref{BZT}, which may involve doubly charged fermions $\tilde{\chi}^{++}$ as well as other particles, can give significant contributions to the muon MDM. After including these two-loop corrections, the muon MDM is given by~\cite{JL2021,ZJY}:
	\begin{eqnarray}
		a_\mu^{\rm two-loop} = a_\mu^{\rm one-loop} + a_\mu^{WW} + a_\mu^{WH} + a_\mu^{\gamma h},
		\label{twoloop MDM}
	\end{eqnarray}
	where $a_\mu^{WW}$, $a_\mu^{WH}$, and $a_\mu^{\gamma h}$ correspond to the contributions from Fig.~\ref{BZT}, with explicit expressions also given in Appendix~\ref{au}.

	\section{The numerical analyses}
	\label{sec4}
	In this section, we present the numerical results for the branching ratios of LFV processes and the MDM of the muon. For the input parameters of the SM, we take the $W$ boson mass $m_W=80.377\;{\rm GeV}$, the $Z$ boson mass $m_Z=90.188\;{\rm GeV}$, the electron mass $m_e=0.511\;{\rm MeV}$, $\alpha_{em}(m_Z)=1/128.9$ for the electromagnetic coupling, the strong coupling $\alpha_s(m_Z)=0.118$, and the top quark mass $m_t = 172.5~\text{GeV}$.
	We impose the cosmological bound \(\sum_i m_{\nu_i} < 0.12\ \mathrm{eV}\)~\cite{PDG,NMTM}.
	In our analysis we assume the charged-lepton mass matrix is diagonal,
	\(M_l=\mathrm{diag}(m_e,m_\mu,m_\tau)\), which implies that the charged-lepton Yukawa matrix \(Y_{h_e}\) is also diagonal.
	This assumption greatly simplifies the determination of the neutrino mixing matrix: the PMNS matrix is then fixed entirely by the neutrino mass matrix~\cite{UPMNS p1,UPMNS p2}.
	Since the absolute neutrino mass ordering is not yet determined, we consider both hierarchies in what follows: the NH with \(m_{\nu_1}<m_{\nu_2}<m_{\nu_3}\) and the IH with \(m_{\nu_3}<m_{\nu_1}<m_{\nu_2}\).
	We use the following \(3\sigma\) ranges for the mass-squared differences and mixing angles~\cite{PDG}:
	
	$$ \Delta m_{\nu_{21}}^2=(6.99-8.07) \times 10^{-5} \;{\rm eV}^2, $$
	$$ | \Delta m_{\nu_{32}}^2 |_\text{NH}=(2.371-2.539)\times 10^{-3} \;{\rm eV}^2 , $$
	$$ | \Delta m_{\nu_{32}}^2 |_\text{IH}=(2.442-2.616)\times 10^{-3} \;{\rm eV}^2 , $$
	$$ \text{sin}^2\theta_{12}=0.270-0.345 , $$
	$$ \text{sin}^2\theta_{23}=0.495-0.603 , $$ 
	\begin{equation}
		\text{sin}^2\theta_{13}=0.0198-0.0240\label{data},
	\end{equation}
	where \( | \Delta m_{\nu_{32}}^2 |_\text{NH} \) corresponds to the NH, and \( | \Delta m_{\nu_{32}}^2 |_\text{IH} \) corresponds to the IH. The Yukawa elements \(Y_{R,ll}\) and \(Y_{D,ll}\) are not independent free parameters: once the diagonal elements $M_{R,ll}\ (l=1,2,3)$ arespecified, Eqs.~(\ref{MLR}) and (\ref{meff}) allow one to express \(Y_{R,ll}\) and \(Y_{D,ll}\) in terms of other parameters.
	Therefore, only the coupling matrix \(Y_L\) associated with the type-II seesaw needs to be specified. In general \(Y_L\) is a \(3\times3\) matrix, which we parametrize as
	\[
	Y_L=\begin{pmatrix}
		Y_{11}&Y_{12}&Y_{13}\\
		Y_{21}&Y_{22}&Y_{23}\\
		Y_{31}&Y_{32}&Y_{33}
	\end{pmatrix}.
	\]
	
	$Y_{11}$ is constrained strongly by the $0\nu2\beta$ decay experiments in the range $Y_{11}\lesssim0.04$ \cite{Yangwz}. For convenience in the subsequent analysis, we fix $Y_{11} = Y_{22} = Y_{33} = 0.04$.
	To simplify the problem, we set the off-diagonal elements to be equal: \( Y_{12} = Y_{21},\, Y_{13} = Y_{31},\, Y_{23} = Y_{32}\). For clarity, we assume that the right-handed neutrino sector is degenerate, 
	with $M_{R,11} = M_{R,22} = M_{R,33} = 500~\mathrm{GeV}$, 
	and that its off-diagonal elements vanish. 
	We note that the right-handed neutrino masses are given by Eq.~\ref{MLR}. 
	Once the parameters $\mu$ and $\lambda$ are fixed, the value of $v_S$ can be determined, 
	which then allows us to obtain the explicit expression for $Y_R$.
	We set  $m_{\nu_{\text{lightest}}}^\text{tree} = 0.01~\mathrm{eV}$ for NH and IH neutrino masses, $s_{12}^2,\,s_{13}^2,s_{23}^2,\,\Delta m_{\nu_{21}}^2,\,|\Delta m_{\nu_{32}}^2|$ with the center values. 
	Once the lightest neutrino mass at tree level, the three mixing angles, 
	and the two mass-squared differences are fixed using neutrino oscillation data, 
	and given that the structure of $Y_L$ and the right-handed Majorana mass matrix $M_R$ is known, 
	as well as the values of $\tan\beta$, $\tan\beta'$, and $v_{T\bar{T}}$, 
	one can determine $v_T$ and $v_u$. 
	Combining this with  Eq.~\ref{MLR} and Eq.~\ref{meff1}, we can then solve for the Dirac Yukawa matrix $Y_D$. In this work, we set all three CP-violating phases of the neutrinos to zero. Including the 
	one-loop radiative corrections, the light neutrino masses are then fully determined. 
	The specific details of these calculations are provided in Appendix~\ref{oloop} for reference.

	The measured mass of the Higgs boson is  \cite{PDG}
	\begin{eqnarray}
		&&m_h=125.25\pm0.17\;{\rm GeV}.
		\label{higgs ma}
	\end{eqnarray}
	
	Taking into account the tree-level relation between the \(W\) and \(Z\) masses, the \(\rho\) parameter in the TNMSSM is~\cite{YX}
	\begin{equation}\label{eq:rho}
		\rho=\frac{m_W^2}{\cos^2\theta_W\, m_Z^2}=1-\frac{2 v_{T\bar{T}}^2}{v_{ud}^2},
	\end{equation}
	where \(\theta_W\) is the Weinberg angle. Current measurements give \(\rho_{\rm exp}=1.00038\pm0.00020\) (at \(2\sigma\)), which implies \(v_{T\bar T}\lesssim 1\ \mathrm{GeV}\) to satisfy electroweak precision constraints.
	Accordingly, in the following we adopt \(v_{T\bar T}=0.001\ \mathrm{GeV}\).

	Current experimental constraints on supersymmetric particle masses are summarized as follows. 
	The lower limit on the mass of long-lived charginos is $m_{\tilde{\chi}^\pm} > 1090~\mathrm{GeV}$ at the 95\% confidence level (CL)~\cite{PDG}. 
	We conservatively take the mass of the lightest neutralino to be larger than $300~\mathrm{GeV}$ by fixing the Higgsino sector parameter to $M_1 = 800~\mathrm{GeV}$ and choosing $M_2 \geq 1000~\mathrm{GeV}$.
	We assume that the soft-breaking mass parameters of the sleptons satisfy 
	$m_{\tilde L}=m_{\tilde e}=m_{\tilde \nu}={\rm diag}(M_E,\,M_E,\,M_E)\;{\rm GeV}$, 
	where $m_{\tilde L}$ and $m_{\tilde e}$ denote the soft-breaking masses of the left-handed slepton doublets and right-handed charged sleptons, respectively, and $m_{\tilde \nu}$ denotes the corresponding sneutrino soft-breaking masses~\cite{Meng1}.
	Current LHC searches exclude slepton masses below about $700~{\rm GeV}$;
	therefore, we take $M_E \gtrsim 800~{\rm GeV}$ in our analysis~\cite{data1,data2,data3}. Further details of the slepton mass matrix, together with the impact of $M_E$ and $A_{he}$ on its structure, are provided in Appendix~\ref{slepton}.
	However, for the trilinear Higgs–slepton couplings $A_{h_e}$, we do not assume a diagonal form. Instead, we allow flavor-mixing off-diagonal terms in the slepton sector, whose explicit structure is given by:
	\begin{eqnarray}
		A_{h_e} & = &
		\left(
		\begin{array}{ccc}
			1 & \delta_{12} & \delta_{13} \\
			\delta_{12} & 1 & \delta_{23} \\
			\delta_{13} & \delta_{23} & 1
		\end{array}
		\right)
		\, A_e \; 
	\end{eqnarray}
	To investigate the variation of the branching ratios in the processes 
	\(\ell_j^- \to \ell_i^-\gamma\) and \(\ell_j^- \to \ell_i^- \ell_i^- \ell_i^+\), 
	which are sensitive to the relevant parameters, we first fix
	several parameters in \(A_{h_e}\). We choose \(A_{e} = 1000~\text{GeV}\).
	
	We take the charged Higgs mass $M_{H^\pm} > 1500~\mathrm{GeV}$~\cite{JL2021}, 
	and the doubly charged Higgs mass is also considered to be $>1500~\mathrm{GeV}$. 
	For squarks, the first two generations are strongly constrained by direct LHC searches~\cite{PDG,Meng1}; therefore we set $m_{\tilde{Q}{1,2}}=m{\tilde{u}^c_{1,2}}=m_{\tilde{d}^c_{1,2}}=3000~\text{GeV}$, while the third generation is less constrained, and we take $m_{\tilde{Q}3}=m{\tilde{u}^c_3}=m_{\tilde{d}^c_3}=2000~\text{GeV}$.
	The trilinear Higgs–squark couplings are defined as
	$A_{hu} \equiv \mathrm{diag}(Y_{u1}, Y_{u2}, Y_{u3}) A_q$ and
	$A_{hd} \equiv \mathrm{diag}(Y_{d1}, Y_{d2}, Y_{d3}) A_q$,
	with $A_q = 1000~\mathrm{GeV}$,
	where $Y_{qi}$ ($q=u,d$, $i=1,2,3$) denote the corresponding Yukawa coupling constants.

	In addition, rare $B$-meson decay processes, such as 
	$\bar{B}\to X_s\gamma$ and $B_s^0\to\mu^+\mu^-$, play an important role in probing 
	new physics (NP) models beyond the SM, since their theoretical predictions are 
	tightly constrained by precise experimental measurements. The current world 
	averages for the branching ratios are given by~\cite{PDG}  
	\begin{eqnarray}
		{\rm Br}(\bar{B}\to X_s\gamma) &=& (3.49 \pm 0.19)\times 10^{-4}, \nonumber \\
		{\rm Br}(B_s^0\to\mu^+\mu^-) &=& (3.01 \pm 0.35)\times 10^{-9}.
	\end{eqnarray}
	Both of these constraints are included in our numerical analysis. For further 
	discussions on these rare decays, see our previous works~\cite{JL2018,ZJY} 
	and references therein.

	The neutrino oscillation experimental data~\cite{PDG} and the mass of the lightest CP-even Higgs boson impose strong constraints on the relevant parameter space. 
	Analytic formulas, including the two-loop leading-log radiative corrections from the stop and top-quark contributions~\cite{HiggsC1,HiggsC2,HiggsC3,HiggsC4}, are used to compute the SM-like Higgs boson mass, as shown in Eq.~\ref{hg2l}.
	For completeness, detailed expressions for the Higgs mass matrix and the two-loop corrections are given in Appendix~\ref{higgs}. 
	Moreover, rare $B$-meson decays—$\bar{B}\to X_s\gamma$ and $B_s^0\to\mu^+\mu^-$—together with the mass bounds on the charged Higgs, doubly charged Higgs, neutralinos, and sleptons, further restrict the allowed parameter space. Special attention is given to the mass of the lightest Higgs boson in Eq.~\ref{higgs ma}. 
	To satisfy all these experimental constraints while reducing the number of free parameters in the numerical analysis, we fix the following parameters:
	
	\begin{eqnarray}
		&& \tan\beta = 4.3, \quad \lambda = 0.5, \quad \lambda_T = 0.22, \quad \kappa = 0.86, \quad \tan\beta' = 0.8, \nonumber\\
		&& \chi_u = 0.1, \quad \chi_d = 1.2, \quad A_{\lambda_T} = A_\lambda = 1000~{\rm GeV}, \quad A_{e} = 1000~{\rm GeV}, \nonumber\\
		&& A_\kappa = -1000~{\rm GeV}, \quad A_{\chi_u} = -850~{\rm GeV}, \quad A_{\chi_d} = -500~{\rm GeV}. \label{para}
	\end{eqnarray}
	Using the \texttt{SARAH~4} package~\cite{sarah} to obtain interaction vertices and particle masses, and taking into account the experimental constraints on the lightest Higgs boson, neutrino oscillation data, rare $B$-meson decays, and the mass bounds on supersymmetric particles, we determine the allowed ranges for the parameters listed in Eq.~\ref{para}. For simplicity, we fix these values throughout the subsequent analysis to focus on the most sensitive parameters, which reduces the number of free parameters while ensuring consistency with neutrino oscillation data, rare $B$-meson decay bounds, and the mass limits on the charged and doubly charged Higgs bosons, neutralinos, and sleptons.
	
	As discussed in Sec.\ref{sec3}, and considering that supersymmetric particle masses are significantly heavier than those of leptons, we adopt an effective Lagrangian approach to study charged LFV processes and the muon $(g-2)_\mu$, with the corresponding analytic expressions summarized in Appendices~\ref{wilsonllr} and~\ref{au} and implemented in a self-developed code for numerical evaluation.
	The resulting predictions for LFV and the muon MDM are analyzed in the following sections.

	\subsection{The Muon MDM}
	\begin{figure}
		\setlength{\unitlength}{1mm}
		\centering
		\includegraphics[width=3in]{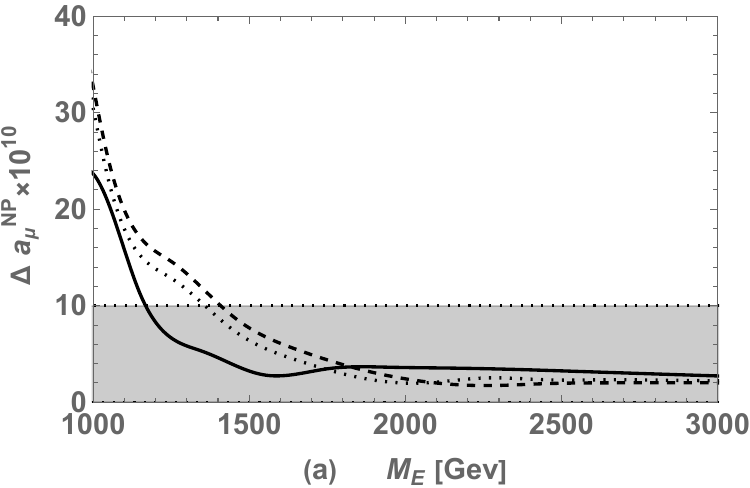}
		\vspace{0cm}
		\includegraphics[width=3in]{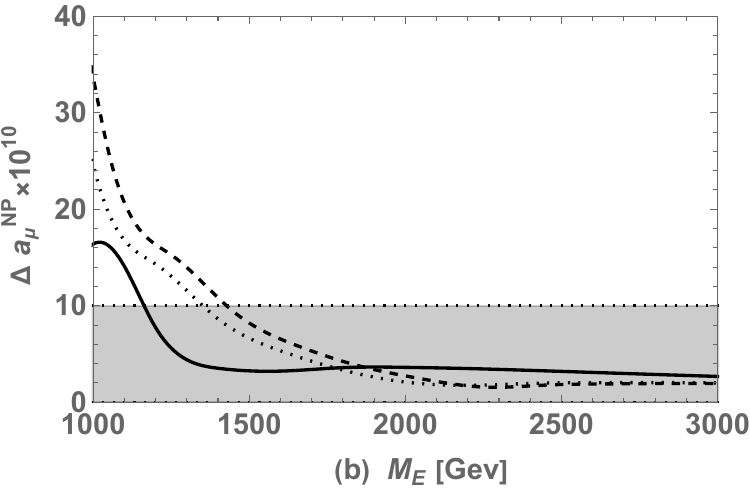}
		\vspace{0cm}
		\caption{The plot shows $\Delta a_\mu^{\mathrm{NP}}$ as a function of $M_E$.
			Panels (a) and (b) correspond to the NH and IH neutrino mass spectra, respectively.
			The solid, dotted, and dashed lines represent the cases where $\mu$ is taken to be 1000~GeV, 1150~GeV, and 1300~GeV, respectively,
			and the gray band denotes the experimental $1\sigma$ interval.
		}
		\label{G2A}
	\end{figure}
	
	We analyze the  MDM of the muon in the TNMSSM framework. According to Eq.~\ref{g2n}, the new physics contribution $\Delta a_\mu^{\rm NP}$ is constrained within $(-2.5 \times 10^{-10},\, 10.3 \times 10^{-10})$ at the $1\sigma$ level. In Fig.~\ref{G2A}, we fix $M_2 = 1000\,\mathrm{GeV}$ and set all off-diagonal elements of the trilinear coupling matrix $A_{h_e}$ and the Yukawa matrix $Y_{LL}$ to zero. Panels~(a) and (b) correspond to NH and IH, respectively, with the line styles indicating $\mu = 1000, 1150, 1300\,\mathrm{GeV}$ (solid, dotted, dashed).  
	
	As $M_E$ increases, $\Delta a_\mu^{\rm NP}$ decreases, initially rapidly and then more gradually. This trend arises because larger $M_E$ values yield heavier sleptons and sneutrinos, suppressing the dominant one-loop contributions. In contrast, two-loop contributions, primarily from charginos and the doubly charged fermion $\chi^{++}$, remain largely unaffected by $M_E$. Consequently, the relative importance of two-loop effects increases, moderating the overall decline. Although smaller in magnitude, these higher-order contributions are non-negligible, indicating that three-loop or even higher-order corrections could be relevant in future precision analyses.

	The parameter $\mu$ also significantly influences $\Delta a_\mu^{\rm NP}$. For small $M_E$, the contribution increases with $\mu$, being smallest at $1000\,\mathrm{GeV}$ and largest at $1300\,\mathrm{GeV}$. As $M_E$ grows, this trend reverses, producing a non-monotonic dependence. This behavior results from the interplay of model parameters: for fixed $\lambda$, increasing $\mu$ enhances the singlet VEV $v_S$, which shifts the masses of charged Higgs bosons, sleptons, and wino-like charginos, thereby modifying loop contributions and interference patterns. As $M_E$ further increases, heavy sleptons and sneutrinos decouple, causing a monotonic suppression of the total loop effects. To remain consistent with current expectations, we fix $M_E = 1500\,\mathrm{GeV}$ in subsequent LFV analyses, also ensuring compatibility with the neutrino oscillation data as discussed in Appendix~\ref{oloop}.

	Finally, the small differences between NH and IH curves reflect the impact of neutrino mass ordering on the Yukawa couplings $Y_L$, $Y_D$, and $Y_R$, which enter the MDM calculation indirectly. While nearly identical, these subtle distinctions suggest that future high-precision measurements of the muon MDM could provide indirect insights into the neutrino mass hierarchy.
	
	\subsection{Branching Ratios for LFV Processes}

	\begin{figure}
		\setlength{\unitlength}{1mm}
		\centering
		\begin{minipage}[c]{0.5\textwidth}
			\includegraphics[width=2.9in]{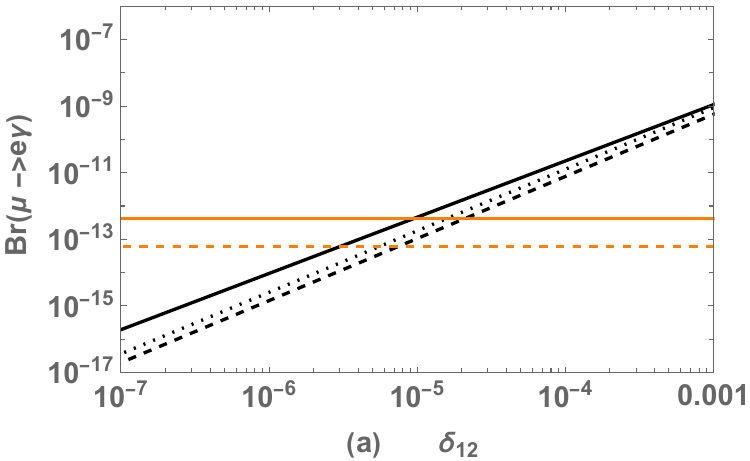}
		\end{minipage}%
		\begin{minipage}[c]{0.5\textwidth}
			\includegraphics[width=2.9in]{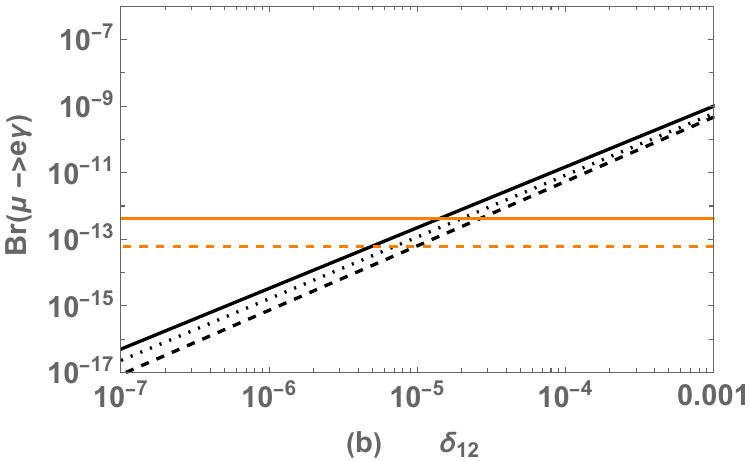}
		\end{minipage}
		\begin{minipage}[c]{0.5\textwidth}
			\includegraphics[width=2.9in]{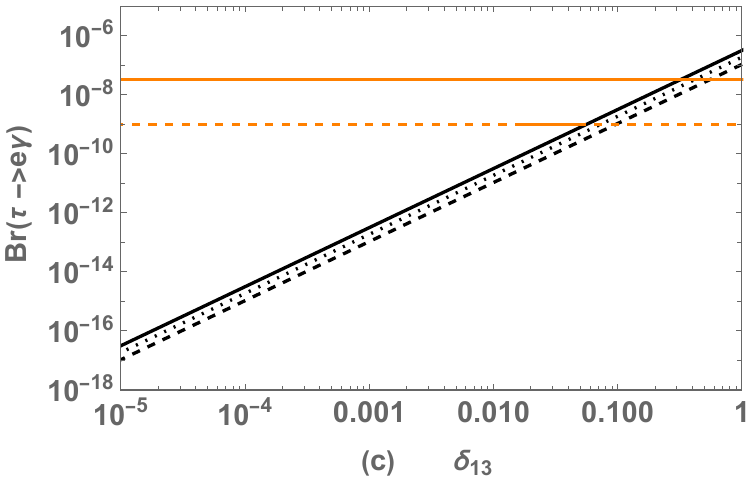}
		\end{minipage}%
		\begin{minipage}[c]{0.5\textwidth}
			\includegraphics[width=2.9in]{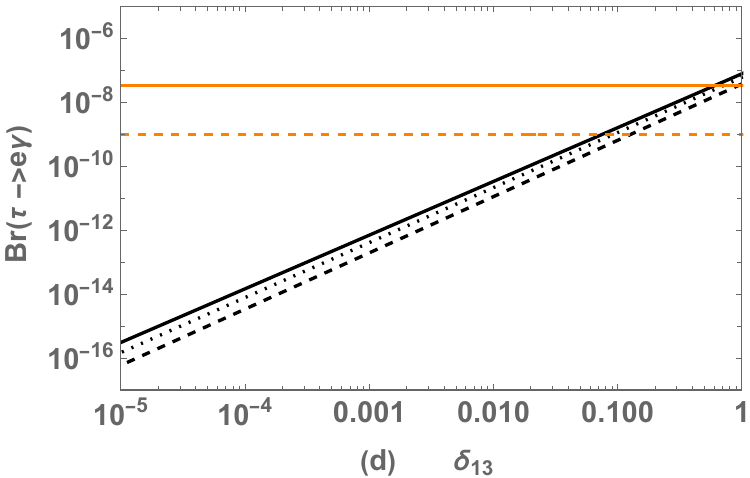}
		\end{minipage}
		\begin{minipage}[c]{0.5\textwidth}
			\includegraphics[width=2.9in]{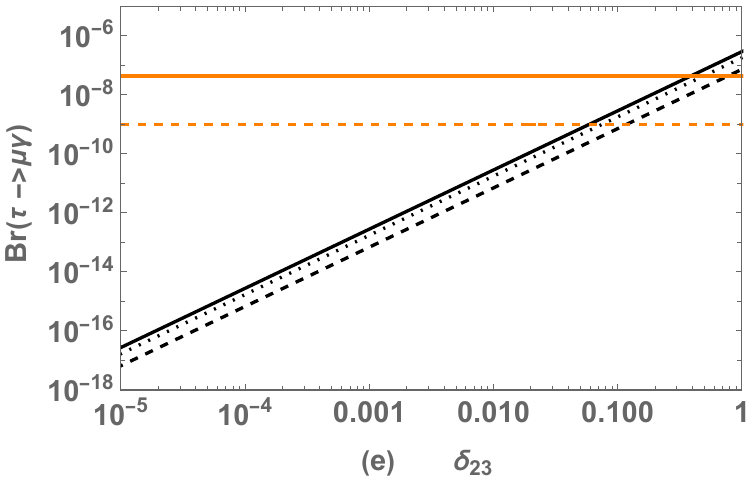}
		\end{minipage}%
		\begin{minipage}[c]{0.5\textwidth}
			\includegraphics[width=2.9in]{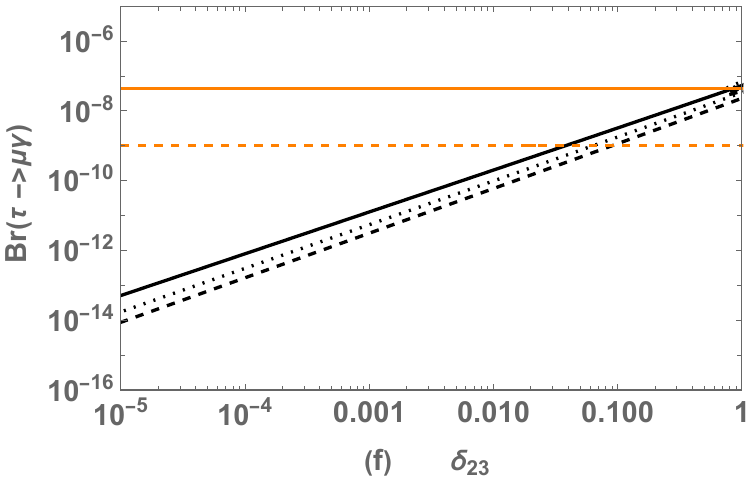}
		\end{minipage}

		\caption{The branching ratio $\mathrm{Br}(\ell_j^- \to \ell_i^- \gamma)$ is shown as a function of the off-diagonal elements $\delta_{ij}$ of the trilinear coupling matrix $A_{h_e}$, with panels (a, c, e) corresponding to the NH and panels (b, d, f) to the IH. 
			The black solid, dotted, and dashed lines represent $M_2 = 1000$, $1250$, and $1500$~GeV, respectively, while the orange solid and dashed lines indicate the current and projected future experimental bounds on LFV.
		}
		\label{NHDET}
	\end{figure}
	
	\begin{figure}
		\setlength{\unitlength}{1mm}
		\centering
		\begin{minipage}[c]{0.5\textwidth}
			\includegraphics[width=2.9in]{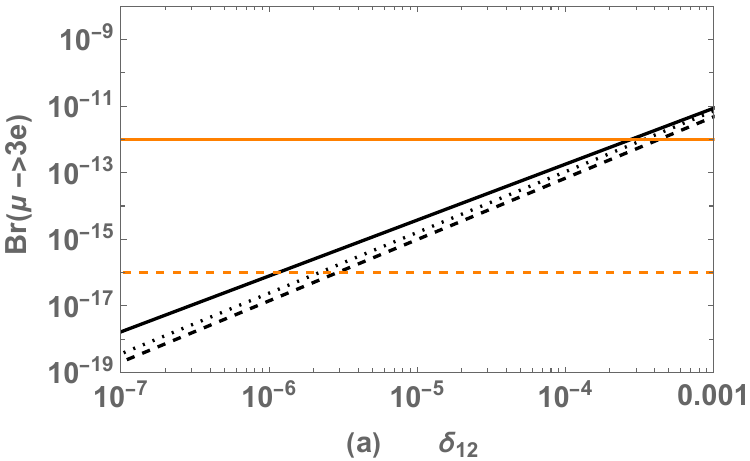}
		\end{minipage}%
		\begin{minipage}[c]{0.5\textwidth}
			\includegraphics[width=2.9in]{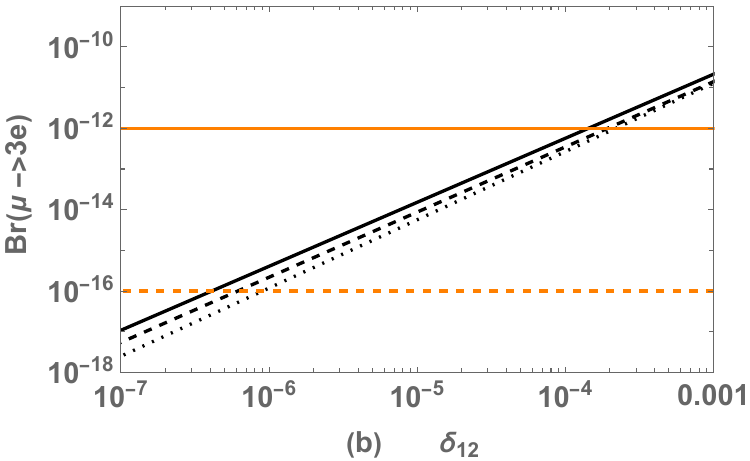}
		\end{minipage}
		\begin{minipage}[c]{0.5\textwidth}
			\includegraphics[width=2.9in]{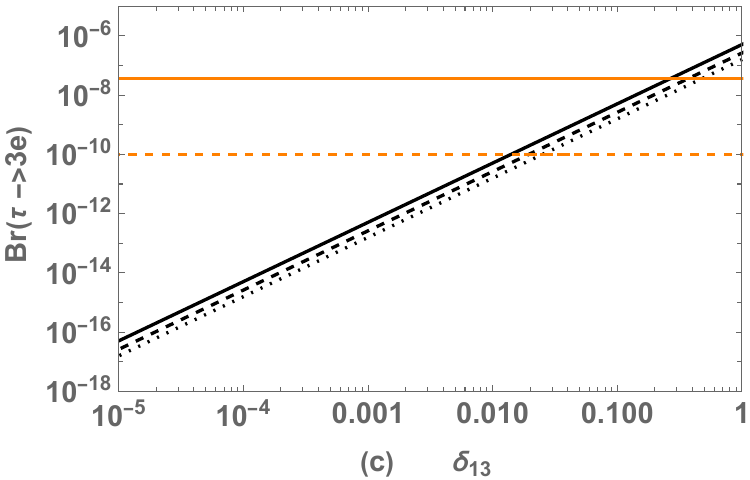}
		\end{minipage}%
		\begin{minipage}[c]{0.5\textwidth}
			\includegraphics[width=2.9in]{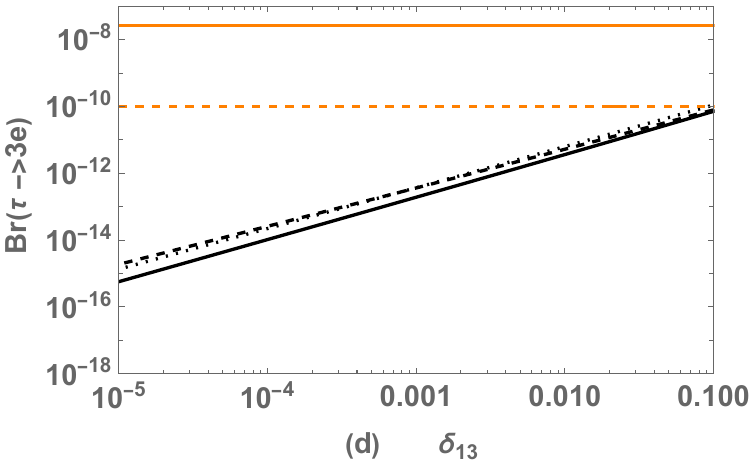}
		\end{minipage}
		\begin{minipage}[c]{0.5\textwidth}
			\includegraphics[width=2.9in]{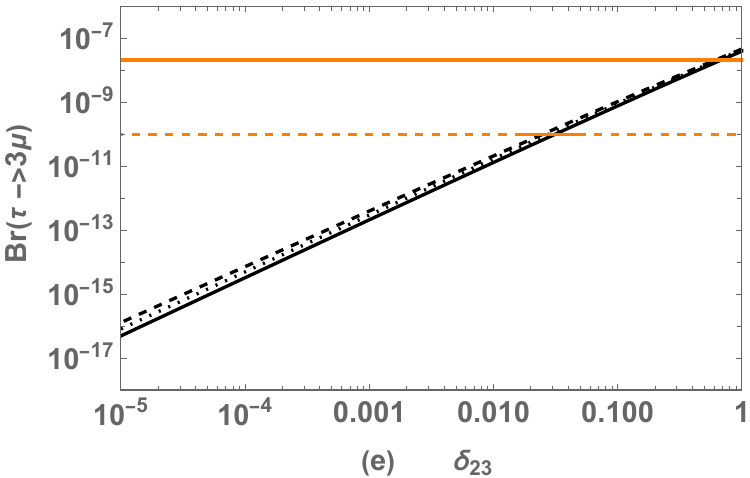}
		\end{minipage}%
		\begin{minipage}[c]{0.5\textwidth}
			\includegraphics[width=2.9in]{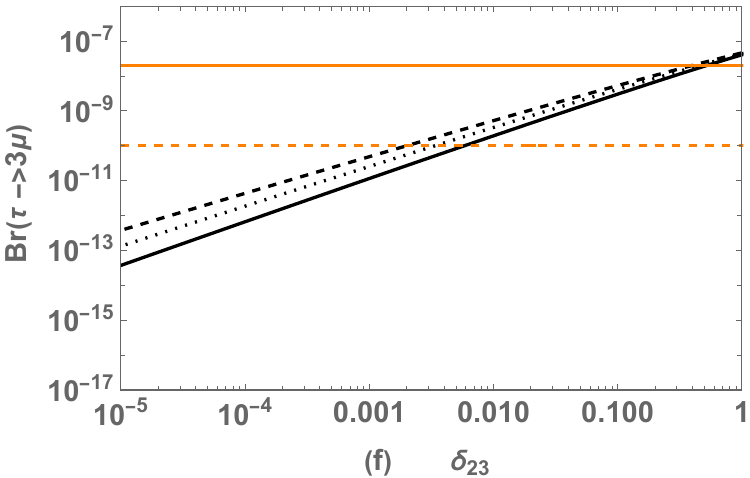}
		\end{minipage}

		\caption{The variation of \(\delta_{ij}\) associated with the LFV branching ratios \(\text{Br}(\ell_j^- \to \ell_i^- \ell_i^- \ell_i^+)\) is shown for both neutrino mass hierarchies: (a), (c), and (e) correspond to the NH, while (b), (d), and (f) correspond to the IH.
			The black solid, dotted, and dashed lines correspond to $M_2 = 1000$, $1250$, and $1500$~GeV, respectively. The orange solid and dashed lines represent the current experimental limits and the projected future sensitivity, respectively.
		}
		\label{IHDET}
	\end{figure}
	In Fig.~\ref{NHDET}, we fix $M_E = 1500\,\mathrm{GeV}$ and $\mu = 1000\,\mathrm{GeV}$ for both NH (panels a, c, e) and IH (panels b, d, f).
	The off-diagonal elements of $Y_L$ are set to zero for $i\neq j$ to ensure flavor alignment. The branching ratios of radiative LFV decays, $\mathrm{Br}(\ell_j^- \to \ell_i^- \gamma)$, are plotted against the off-diagonal parameters $\delta_{ij}$ of $A_{h_e}$. Panels (a,b) show $\mathrm{Br}(\mu \to e \gamma)$ versus $\delta_{12}$ with $\delta_{13} = \delta_{23} = 0$; panels (c,d) present $\mathrm{Br}(\tau \to e \gamma)$ versus $\delta_{13}$; and panels (e,f) display $\mathrm{Br}(\tau \to \mu \gamma)$ versus $\delta_{23}$. Solid, dashed, and dotted lines correspond to $M_2 = 1000$, $1250$, and $1500\,\mathrm{GeV}$, respectively, while the orange solid and dashed lines represent the current experimental bounds and the projected future sensitivities, respectively.
	From Fig.~\ref{NHDET}, the experimental bound on $\mathrm{Br}(\mu \to e \gamma)$ implies $\delta_{12} \lesssim 10^{-5}$, whereas $\tau \to e \gamma$ and $\tau \to \mu \gamma$ require $\delta_{13}, \delta_{23} \lesssim 0.5$. Differences between NH and IH arise because, for fixed $Y_L$ and $M_R$, the Dirac Yukawa couplings $Y_D$ must adjust to reproduce the neutrino mass hierarchy, altering SUSY particle mixing and LFV rates.
	
	In Fig.~\ref{IHDET}, the same setup is applied to three-body LFV decays, $\mathrm{Br}(\ell_j^- \to \ell_i^- \ell_i^- \ell_i^+)$. Panels (a,b) show $\mathrm{Br}(\mu \to 3e)$ versus $\delta_{12}$; panels (c,d) show $\mathrm{Br}(\tau \to 3e)$ versus $\delta_{13}$; and panels (e,f) display $\mathrm{Br}(\tau \to 3\mu)$ versus $\delta_{23}$. The line styles follow the same $M_2$ convention. The branching ratios increase with $\delta_{ij}$ because of enhanced LFV. Current bounds suggest $\delta_{12} \lesssim 2\times10^{-4}$, $\delta_{13} \lesssim 0.1$, and $\delta_{23} \lesssim 0.7$. Although $M_2$ influences the three-body decays, the effect is weaker than in radiative decays, since these processes are dominated by tree-level contributions in the TNMSSM. The mild non-monotonic $M_2$ dependence originates from partial cancellations among loop diagrams.

	\begin{figure}
		\setlength{\unitlength}{1mm}
		\centering
		\begin{minipage}[c]{0.5\textwidth}
			\includegraphics[width=2.9in]{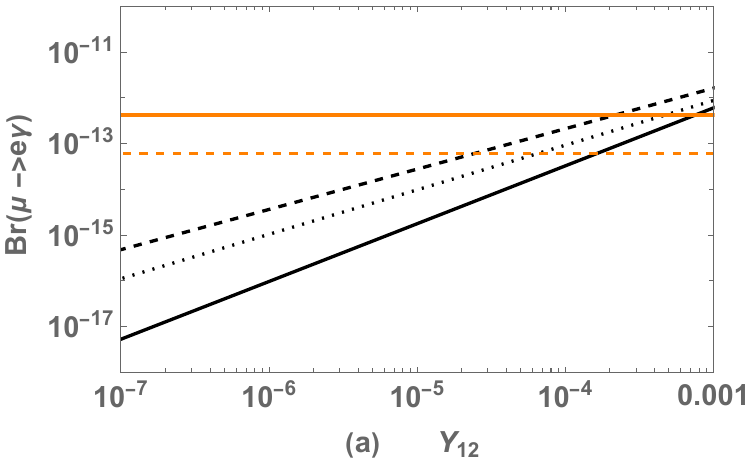}
		\end{minipage}%
		\begin{minipage}[c]{0.5\textwidth}
			\includegraphics[width=2.9in]{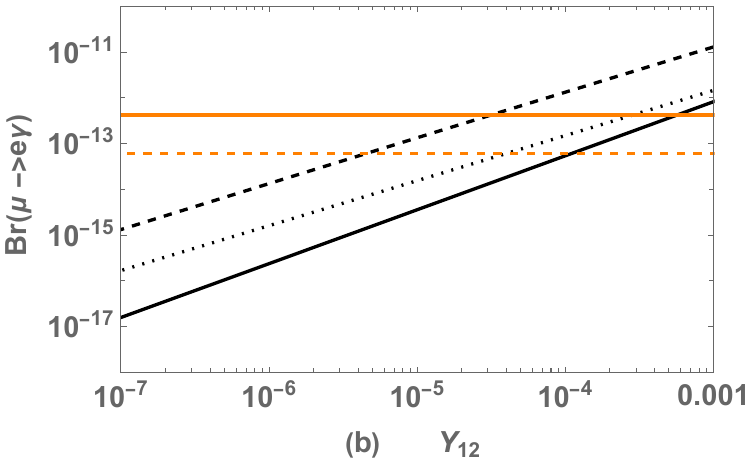}
		\end{minipage}
		\begin{minipage}[c]{0.5\textwidth}
			\includegraphics[width=2.9in]{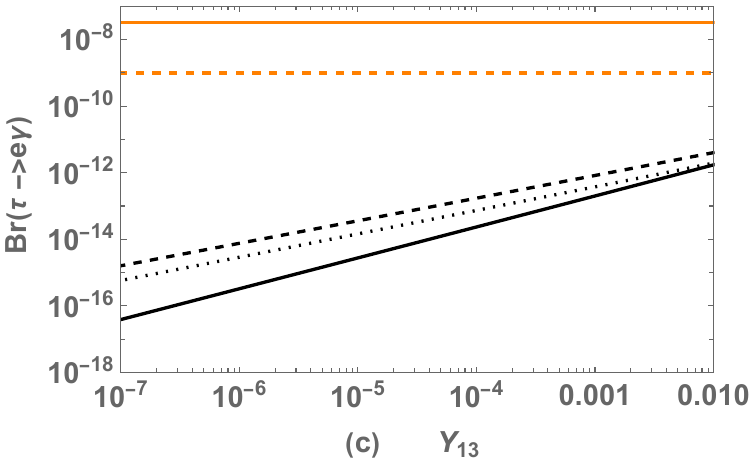}
		\end{minipage}%
		\begin{minipage}[c]{0.5\textwidth}
			\includegraphics[width=2.9in]{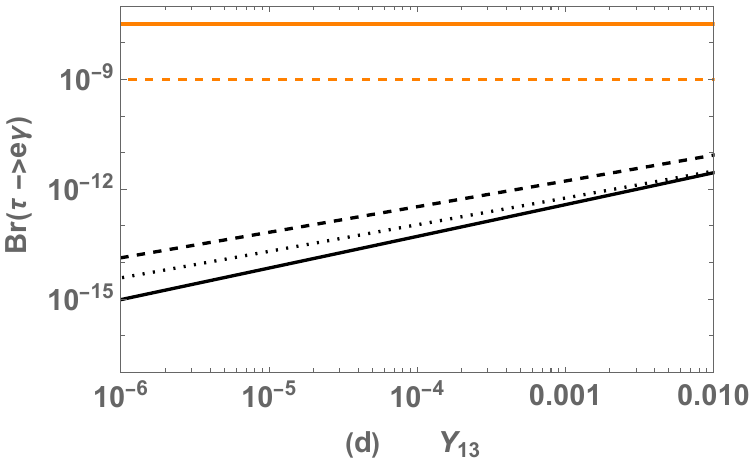}
		\end{minipage}
		\begin{minipage}[c]{0.5\textwidth}
			\includegraphics[width=2.9in]{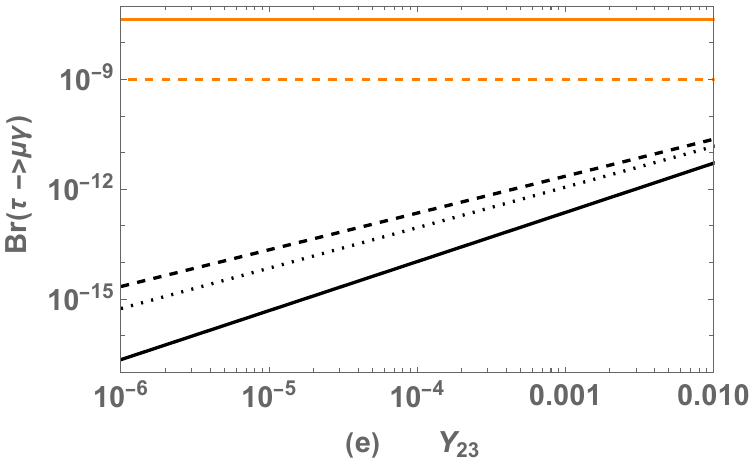}
		\end{minipage}%
		\begin{minipage}[c]{0.5\textwidth}
			\includegraphics[width=2.9in]{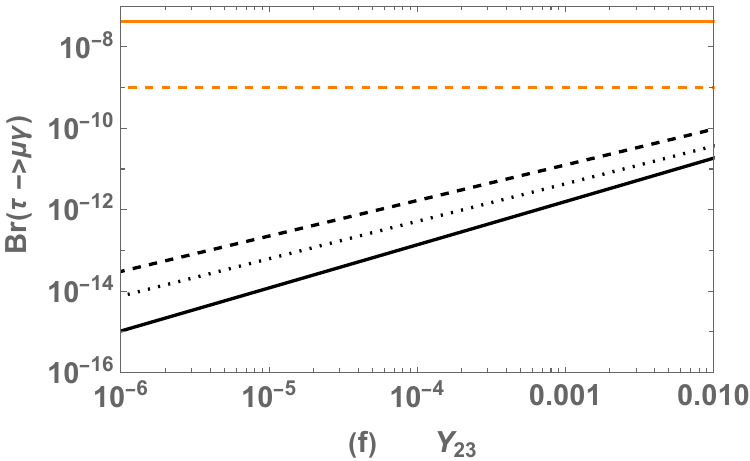}
		\end{minipage}

		\caption{The dependence of \(Y_{ij}\) on the LFV branching ratios 
			\(\mathrm{Br}(\ell_j^- \to \ell_i^-\gamma)\) is shown for both neutrino mass
			hierarchies: panels (a), (c), and (e) correspond to the NH,
			while (b), (d), and (f) display the IH. 
			In (a)–(b), the solid, dotted, and dashed lines indicate 
			\(\delta_{12} = 10^{-7}, 5 \times 10^{-7}, 10^{-6}\), respectively; 
			in (c)–(d) they indicate 
			\(\delta_{13} = 10^{-5}, 5 \times 10^{-5}, 10^{-4}\); 
			and in (e)–(f) the same line styles correspond to 
			\(\delta_{23}\) with the same set of values. 
			The orange solid and dashed lines represent the current experimental limits and
			the projected future sensitivity, respectively.
		}
		\label{YLLLF}
		
	\end{figure}
	The following discussion focuses on the role of the Yukawa couplings \(Y_{L_{ij}}\) within the Type-II seesaw framework. Figure~\ref{YLLLF} shows the branching ratios of radiative LFV decays, \(\mathrm{Br}(\ell_j^- \to \ell_i^-\gamma)\), as functions of the off-diagonal Yukawa elements \(Y_{L_{ij}}\) (\(i\neq j\)). We fix \(M_E = 1500\,\mathrm{GeV}\), \(M_2 = 1000\,\mathrm{GeV}\), and \(\mu = 1000\,\mathrm{GeV}\). Panels (a, c, e) correspond to the NH, while (b, d, f) correspond to the IH. In panels (a,b), the solid, dotted, and dashed lines denote \(\delta_{12} = 10^{-7}, 5\times10^{-7}, 10^{-6}\), respectively; in panels (c–d), they denote \(\delta_{13} = 10^{-5}, 5\times10^{-5}, 10^{-4}\); 
	in panels (e–f), they denote \(\delta_{23} = 10^{-5}, 5\times10^{-5}, 10^{-4}\). Orange lines indicate current experimental limits (solid) and projected future sensitivities (dashed).
	According to Eq.~\ref{eq:lltl}, one can see that even if the off-diagonal elements of \(Y_L\) are non-zero, the LFV process \(\ell_j^- \to \ell_i^-\gamma\) receives no tree-level contribution. Therefore, the amplitude for \(\mathrm{Br}(\ell_j^- \to \ell_i^-\gamma)\) comes entirely from one-loop diagrams.
	Increasing the off-diagonal Yukawa elements \(Y_{L_{ij}}\) requires larger Dirac Yukawa elements \(Y_D\) to satisfy neutrino mass constraints. Both \(Y_L\) and \(Y_D\) enhance mixings between sneutrinos and charginos, and between charged leptons and doubly-charged Higgs bosons, thereby increasing the loop amplitude. However, the overall branching ratio remains moderate due to the smallness of these loop contributions. As a result, current bounds constrain \(Y_{12} \lesssim 10^{-4}\) for \(\delta_{12} \sim 10^{-7}\), providing a quantitative measure of the allowed off-diagonal Yukawa coupling.

	\begin{figure}
		\setlength{\unitlength}{1mm}
		\centering
		\begin{minipage}[c]{0.5\textwidth}
			\includegraphics[width=2.8in]{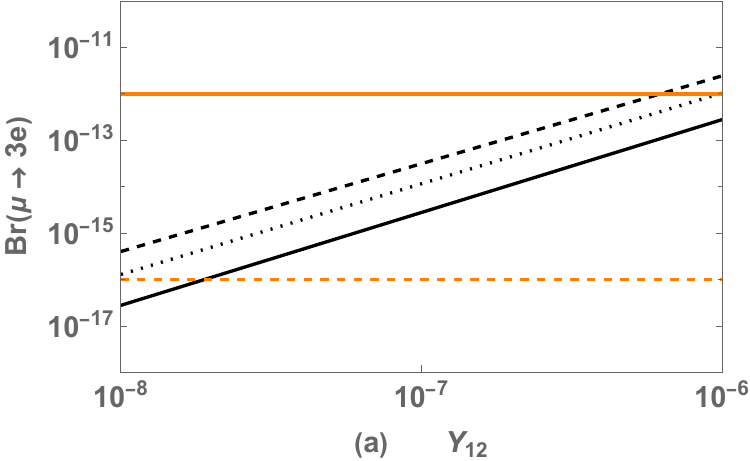}
		\end{minipage}%
		\begin{minipage}[c]{0.5\textwidth}
			\includegraphics[width=2.8in]{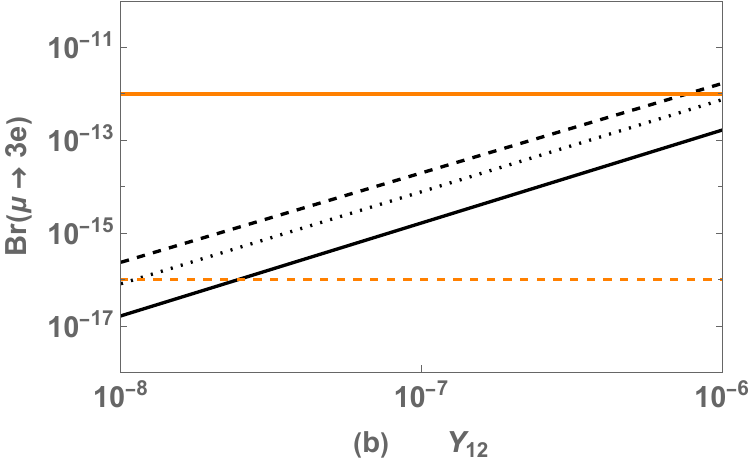}
		\end{minipage}
		\begin{minipage}[c]{0.5\textwidth}
			\includegraphics[width=2.8in]{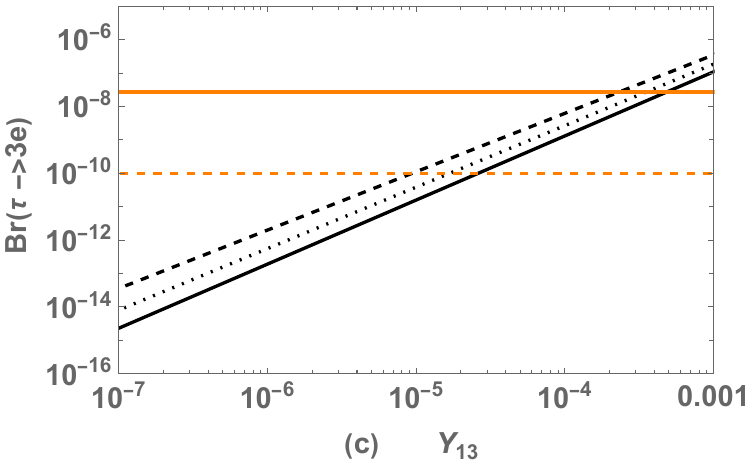}
		\end{minipage}%
		\begin{minipage}[c]{0.5\textwidth}
			\includegraphics[width=2.8in]{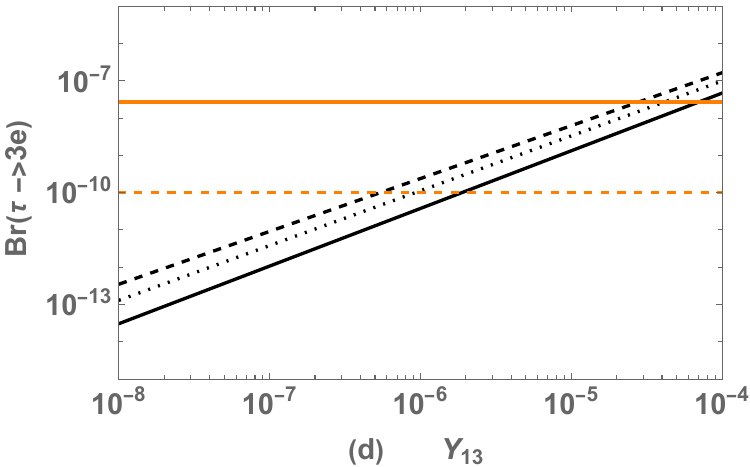}
		\end{minipage}
		\begin{minipage}[c]{0.5\textwidth}
			\includegraphics[width=2.8in]{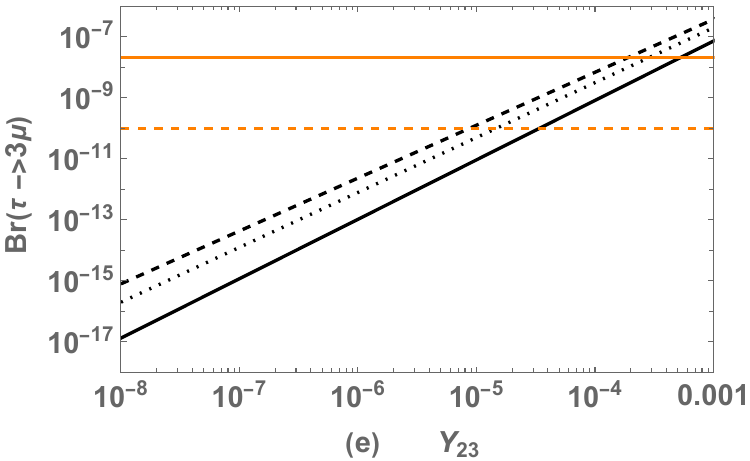}
		\end{minipage}%
		\begin{minipage}[c]{0.5\textwidth}
			\includegraphics[width=2.8in]{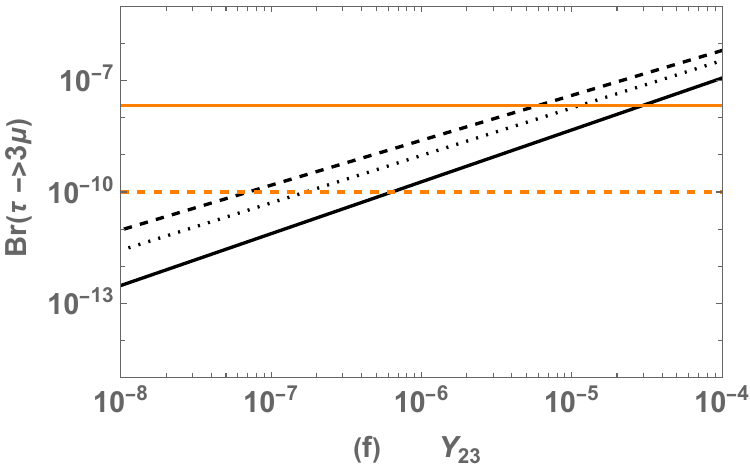}
		\end{minipage}	
		\caption{The variation of \( Y_{L_{ij}} \) associated with the LFV branching ratios \(\text{Br}(\ell_j^- \to \ell_i^- \ell_i^- \ell_i^+)\) is shown for both neutrino mass hierarchies: plots (a), (c), and (e) correspond to the NH, while (b), (d), and (f) correspond to the IH. In panels (a)–(b), solid, dotted, and dashed lines denote \(\delta_{12} = 10^{-7}, 5 \times 10^{-7}, 10^{-6}\), respectively; in panels (c)–(d), they denote \(\delta_{13} = 10^{-5}, 5 \times 10^{-5}, 10^{-4}\); and in panels (e)–(f), \(\delta_{23}\) takes the same values. The orange solid and dashed lines indicate the current experimental limits and the projected future sensitivities, respectively.
		}
		\label{YLLF}
	\end{figure}
	
	Figure~\ref{YLLF} shows three-body LFV decays, \(\text{Br}(\ell_j^- \to \ell_i^- \ell_i^- \ell_i^+)\), as functions of \(Y_{L_{ij}}\), with the same parameter setup. Panels (a, c, e) correspond to NH, while (b, d, f) correspond to IH. The line conventions follow those in Fig.~\ref{YLLLF}. Unlike radiative decays, these processes receive tree-level contributions from doubly-charged Higgs exchange, making them more sensitive to \(Y_L\). The branching ratios increase significantly as $Y_{ij}$ grows.
	The convergence of the curves at large \(Y_{ij}\) indicates that \(\delta_{ij}\)-induced effects become less important compared to loop corrections.
	Considering current experimental limits, we find \(Y_{12} \lesssim 7\times10^{-7}\) (panels a,b), \(Y_{13} \lesssim 7\times10^{-5}\) (c,d), and \(Y_{23} \lesssim 10^{-4}\) (e,f). The branching ratios also show mild differences between NH and IH, reflecting the hierarchy-dependent structure of the Yukawa couplings and the resulting mixings among charged leptons.
	
	Currently, the neutrino mass ordering, CP-violating phases, and the Majorana or Dirac nature of neutrinos remain unknown.  Future theoretical and experimental studies of charged LFV decays could provide valuable insights into these fundamental properties, underlining the importance of this research direction in particle physics.

	\section{Conclusions}
	\label{sec5}
	This paper investigates LFV processes and the two-loop contributions to the muon  MDM, \( \Delta a_\mu^{\mathrm{NP}} \), within the framework of the TNMSSM. In this model, neutrino masses are generated via a combination of Type-I and Type-II seesaw mechanisms. Additional one-loop radiative corrections are included in the analysis.
	
	Under the constraints of current experimental data in neutrino physics, we explore the implications of two neutrino mass orderings: the NH and the IH. Compared to the MSSM, the TNMSSM features an extended particle spectrum—including additional sleptons, sneutrinos, neutralinos, Higgs bosons, as well as doubly charged Higgs bosons and their supersymmetric partners—which significantly enriches the phenomenology and opens new channels for LFV processes.

	In particular, the doubly charged Higgs bosons introduced by the Type-II seesaw mechanism can mediate LFV decays such as $\ell_j^- \to \ell_i^- \ell_i^- \ell_i^+$ at tree level. This enhances their branching ratios.
	Our numerical analysis shows that the branching ratios \(\mathrm{Br}(\ell_j^- \to \ell_i^-\gamma)\) and \(\text{Br}(\ell_j^- \to \ell_i^- \ell_i^- \ell_i^+)\) increase with the enhancement of the off-diagonal elements in the trilinear coupling matrix $A_{h_e}$.
	Moreover, loop diagrams involving neutralinos and charginos also contribute significantly to these processes.

	The predictions exhibit sensitivity to the neutrino mass hierarchy, and if future neutrino oscillation experiments cannot resolve the ordering, LFV decays may serve as an alternative probe, potentially providing complementary insights into the fundamental nature of neutrinos and helping to distinguish between Majorana and Dirac neutrinos.
	
	We have further analyzed the muon MDM in the allowed parameter space of the TNMSSM. When two-loop contributions are included, the doubly charged fermion $\chi^{++}$ plays an essential role. As the soft mass parameter $M_E$ increases, the masses of the associated sleptons and sneutrinos grow, suppressing the one-loop contributions, while the two-loop contributions become increasingly relevant. The combined one- and two-loop effects yield predictions for $\Delta a_\mu^{\mathrm{NP}}$ that are consistent with the latest experimental results, demonstrating the significant impact of the extended particle spectrum on the muon MDM.
	
	With the continuous advancement in lepton physics, the uncertainty in the new physics contribution to the muon MDM, $\Delta a_\mu^{\mathrm{NP}}$, has been substantially reduced. Recent studies indicate that the observed anomaly could be explained within the SM; nevertheless, this does not completely exclude potential contributions from new physics. As experimental precision continues to improve, including higher-order corrections—such as two-loop or even three-loop contributions—may be necessary to maintain consistency between theoretical predictions and measurements. This highlights the ongoing importance of precision studies in probing physics beyond the SM.

	Although the TNMSSM represents a relatively minimal extension of the MSSM, it offers a considerably richer phenomenology. It can enhance new physics contributions to both LFV processes and the muon MDM to potentially observable levels. Current theoretical and experimental studies of the muon MDM may indirectly shed light on undetected LFV phenomena and supersymmetric physics. While the TNMSSM is not necessarily a final theory, it provides a compelling framework that deepens our understanding of particle physics and may guide future efforts to resolve outstanding questions in the field.

	\section*{Acknowledgements}
	The work has been supported by Natural Science Foundation of Guangxi Autonomous
	Region with Grant No. 2022GXNSFDA035068, the National Natural Science Foundation
	of China (NNSFC) with Grants No. 12075074, No. 12235008, No. 11535002, and No.
	11705045, Hebei Natural Science Foundation for Distinguished Young Scholars with Grant
	No. A2022201017, and the youth top-notch talent support program of the Hebei Province.
	
	\appendix
	\section{The one-loop corrections to the neutrino masses\label{oloop}}	
	
	In this section we present the one-loop corrections to neutrino masses in the TNMSSM, following an on-shell renormalization procedure but streamlining the derivation for clarity (see also Ref.~\cite{ZJ1} for related treatments).  
	The general self-energy for neutral–neutrino mixing ${\nu_i}^0\!-\!{\nu_j}^0$ takes the form~\cite{onel2}  
	
	\begin{eqnarray}
		&&\Sigma(k)_{ij}=c_{ij}m_{j}\omega_-+d_{ij}m_{i}\omega_+
		+e_{ij}/\!\!\! k\omega_-+f_{ij}/\!\!\! k \omega_+ .
		\label{onel1}
	\end{eqnarray}
	
	For $k^2\ll m_0^2$—with $m_0$ the heaviest internal mass—the coefficients can be expanded as~\cite{yao}
	\begin{eqnarray}
		&&c_{ij}=c_{ij}^0+k^2c_{ij}^1,\quad 
		d_{ij}=d_{ij}^0+k^2d_{ij}^1,\quad 
		e_{ij}=e_{ij}^0+k^2e_{ij}^1,\quad 
		f_{ij}=f_{ij}^0+k^2f_{ij}^1 .
		\label{series}
	\end{eqnarray}
	
	Counterterms are introduced,
	\begin{equation}
		\Sigma_{ij}^\text{Ren}(k)=\Sigma_{ij}(k)+\big(c_{ij}^*m_{j}\omega_-+d_{ij}^*m_i
		\omega_++e_{ij}^*/\!\!\!k\omega_-+f_{ij}^*/\!\!\!k\omega_+\big),
		\label{counter}
	\end{equation}
	and the on-shell conditions
	\begin{eqnarray}
		\Sigma^\text{Ren}_{ij}(k)u_i(k)\big|_{k^2=m_i^2}=0,\qquad
		\bar{u}_{j}(k)\Sigma^\text{Ren}_{ij}(k)\big|_{k^2=m_j^2}=0 ,
		\label{shell}
	\end{eqnarray}
	yield
	\begin{eqnarray}
		&&c_{ij}^*=-c_{ij}^0+m_i^2d_{ij}^1+m_i^2e_{ij}^1+m_im_jf_{ij}^1,\nonumber \\
		&&d_{ij}^*=-d_{ij}^0+m_j^2c_{ij}^1+m_j^2e_{ij}^1+m_im_jf_{ij}^1,\nonumber \\
		&&e_{ij}^*=-e_{ij}^0-m_j^2c_{ij}^1-m_i^2d_{ij}^1-(m_i^2+m_j^2)e_{ij}^1
		-m_im_jf_{ij}^1,\nonumber \\
		&&f_{ij}^*=-f_{ij}^0-m_im_jc_{ij}^1-m_im_jd_{ij}^1-m_im_je_{ij}^1
		-(m_i^2+m_j^2)f_{ij}^1 .
		\label{solu1}
	\end{eqnarray}
	
	Substituting into Eq.~(\ref{counter}) gives
	\begin{eqnarray}
		&&\Sigma^\text{Ren}_{ij}(k)
		=(/\!\!\!k-m_j)\hat{\Sigma}_{ij}(k)(/\!\!\!k-m_i),
		\label{solu2}
	\end{eqnarray}
	with
	\begin{eqnarray}
		&&\hat{\Sigma}_{ij}(k)=c_{ij}^1m_j\omega_{+}+d_{ij}^1m_i\omega_-
		+e_{ij}^1(m_i\omega_-+m_j\omega_++/\!\!\!k\omega_+)\nonumber \\
		&&\hspace{1.5cm}+f_{ij}^1(m_i\omega_++m_j\omega_-+/\!\!\!k\omega_-).
		\label{hatsig}
	\end{eqnarray}
	
	For convenience define
	\begin{eqnarray}
		&&\delta Z_{ij}^{L}=-m_j^2c_{ij}^1-m_i^2d_{ij}^1-(m_i^2+m_j^2)e_{ij}^1
		-m_im_jf_{ij}^1+e_{ij}^1k^2,\nonumber \\
		&&\delta Z_{ij}^{R}=-m_im_jc_{ij}^1-m_im_jd_{ij}^1-m_im_je_{ij}^1
		-(m_i^2+m_j^2)f_{ij}^1+f_{ij}^1k^2,\nonumber \\
		&&\delta m_{ij}^{L}=\big(m_i^2d_{ij}^1+m_i^2e_{ij}^1+m_im_jf_{ij}^1
		+c_{ij}^1k^2\big)m_j,\nonumber \\
		&&\delta m_{ij}^{R}=\big(m_j^2c_{ij}^1+m_j^2e_{ij}^1+m_im_jf_{ij}^1
		+d_{ij}^1k^2\big)m_i .
		\label{symbel}
	\end{eqnarray}
	
	The two-point Green function up to one loop is then
	\begin{eqnarray}
		&&\Gamma_{ij}(k)=\Big(/\!\!\!k-m_i^\text{{tree}}\Big)\delta_{ij}+\Sigma_{ij}^\text{{Ren}}(k)
		\nonumber \\
		&&\hspace{1.0cm}=\Big(/\!\!\!k-m_i^\text{{tree}}\Big)\delta_{ij}+\delta Z_{ij}^{L}
		/\!\!\!k\omega_-+\delta Z_{ij}^R/\!\!\!k\omega_+
		-\delta m_{ij}^L\omega_--\delta m_{ij}^R\omega_+
		\nonumber \\
		&&\hspace{1.0cm}=(\delta_{ij}+\delta Z_{ij}^L)\Big(/\!\!\!k-m_i^\text{{tree}}
		-\delta m_{ij}^L+\delta Z_{ij}^Lm_i^\text{{tree}}\Big)\omega_- \nonumber \\
		&&\hspace{1.5cm}+(\delta_{ij}+\delta Z_{ij}^R)\Big(/\!\!\!k-m_i^\text{{tree}}
		-\delta m_{ij}^R+\delta Z_{ij}^Rm_j^\text{{tree}}\Big)\omega_+,
		\label{twopoint}
	\end{eqnarray}
	where $\delta Z_{ij}^{L,R}$ act as left/right wave-function renormalization factors and $m_i^{\text{tree}}$ is the tree-level mass.  
	Applying the on-shell conditions finally yields
	\begin{eqnarray}
		&&\delta m_{ij}^\text{loop}=\Big\{[\delta m_{ij}^L+\delta m_{ij}^R]_{k^2=0}
		-\big[m_i\delta Z_{ij}^L\big|_{k^2=m_i^2}
		+m_j\delta Z_{ij}^R\big|_{k^2=m_j^2}\big]\Big\}\nonumber \\
		&&\hspace{1.0cm}=3m_i^{\text{tree}}(m_j^{\text{tree}})^2c_{ij}^1
		+(m_i^{\text{tree}}m_j^{\text{tree}}
		+(m_i^{\text{tree}})^2+(m_j^{\text{tree}})^2)m_i^{\text{tree}}d_{ij}^1
		\nonumber \\
		&&\hspace{1.5cm}+((m_i^{\text{tree}})^2m_j^{\text{tree}}
		+3m_i^{\text{tree}}(m_j^{\text{tree}})^2)e_{ij}^1
		+(3(m_i^{\text{tree}})^2m_j^{\text{tree}}
		+m_i^{\text{tree}}(m_j^{\text{tree}})^2)f_{ij}^1 ,
		\label{loopmass}
	\end{eqnarray}
	which is the final one-loop correction to the neutrino mass matrix.

	\begin{figure}
		\setlength{\unitlength}{1mm}
		\centering
		\begin{minipage}[c]{0.5\textwidth}
			\includegraphics[width=2.9in]{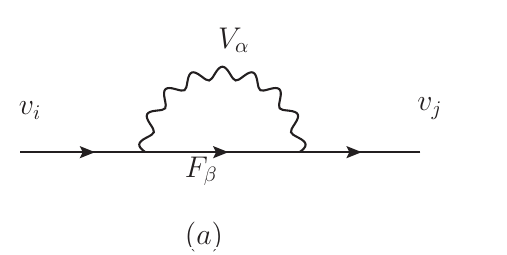}
		\end{minipage}%
		\begin{minipage}[c]{0.5\textwidth}
			\includegraphics[width=2.9in]{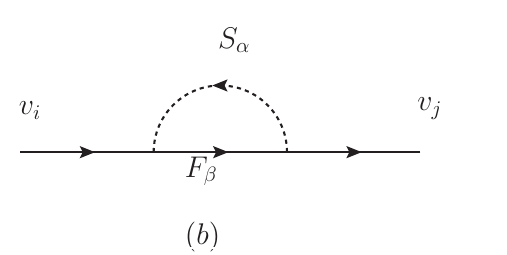}
		\end{minipage}
		\caption{One-loop neutrino self-energy corrections in the TNMSSM from fermion–vector (\(F_\beta\), \(V_\alpha\)) and fermion--scalar (\(F_\beta\), \(S_\alpha\)) loops.
		}
		\label{OF1}
	\end{figure}

	Fig.~\ref{OF1} shows the one-loop self-energy Feynman diagrams for neutrinos in the TNMSSM. The exchanged bosons can be either vector or scalar particles, leading to distinct loop integrals. Specifically, Fig.~\ref{OF1}~(a) corresponds to loops with fermion–vector boson pairs (\(F_\beta, V_\alpha\)), while Fig.~\ref{OF1}~(b) involves fermion–scalar boson pairs (\(F_\beta, S_\alpha\)).
	For the contributions from vector boson loops, the one-loop amplitude can be written as
	\begin{eqnarray}
		&&\text{Amp}_{V}(k)=(\mu_{w})^{2\epsilon}\int\frac{d^DQ}{(2\pi)^D}(iA_{\sigma_1}^{(
			{\tiny V})}\gamma_{\mu}\omega_{\sigma_1})\frac{i(/\!\!\!Q+/\!\!\!k+m_f)}
		{(Q+k)^2-m_f^2}(iB_{\sigma_2}^{({\tiny V})}\gamma^{\mu}\omega_{\sigma_2})
		\frac{-i}{Q^2-m_{{\tiny V}}^2}\nonumber \\
		&&\hspace{1.5cm}=-\int_0^1dx\int\frac{d^DQ}{(2\pi)^D}\frac{1}{(Q^2+x(1-x)k^2
			-xm_f^2-(1-x)m_{\tiny V}^2)^2}\nonumber \\
		&&\hspace{3.0cm}\Big\{(2-D)A_{\sigma}^{({\tiny V})}
		B_{\sigma}^{({\tiny V})}(1-x)/\!\!\!k\omega_{\sigma}
		+Dm_f
		A_{\bar{\sigma}}^{({\tiny V})}B_{\sigma}^{({\tiny V})}\omega_{\sigma}\Big\}
		\nonumber \\
		&&\hspace{1.5cm}=-i\int_0^1dx\int\frac{d^DQ}{(2\pi)^D}\frac{1}{(Q^2
			+xm_f^2+(1-x)m_{\tiny V}^2)^2}\Big\{1+\frac{2x(1-x)k^2}{Q^2
			+xm_f^2+(1-x)m_{\tiny V}^2}\Big\}\nonumber \\
		&&\hspace{2.0cm}\Big\{(2-D)A_{\sigma}^{({\tiny V})}
		B_{\sigma}^{({\tiny V})}(1-x)/\!\!\!k\omega_{\sigma}+Dm_f
		A_{\bar{\sigma}}^{({\tiny V})}B_{\sigma}^{({\tiny V})}\omega_{\sigma}\Big\},
		\label{ampv1}
	\end{eqnarray}
	where $D=4-2\epsilon$ and $\mu_w$ is the renormalization scale.  
	The couplings $A_{\sigma}^{({\tiny V})}$ and $B_{\sigma}^{({\tiny V})}$ with $\sigma=\pm$ denote the interaction vertices, which can be obtained using SARAH~\cite{sarah}.  
	Here $m_{\tiny V}$ and $m_f$ represent the masses of the vector boson and fermion circulating in the loop, respectively.
	
	Using Eq.(\ref{onel1}), Eq.(\ref{series})
	and Eq.(\ref{ampv1}), the coefficients for the self-energy expansion become
	\begin{eqnarray}
		&&c_{ij}^0(m_{\tiny V},m_f)=-iD\frac{m_f}{m_j}A_+^{({\tiny V})}B_-^{({\tiny V})}F_{2a}(m_f,m_{\tiny V}), \quad
		d_{ij}^0(m_{\tiny V},m_f)=-iD\frac{m_f}{m_i}A_-^{({\tiny V})}B_+^{({\tiny V})}F_{2a}(m_f,m_{\tiny V}), \nonumber\\
		&&e_{ij}^0(m_{\tiny V},m_f)=-i(2-D)A_-^{({\tiny V})}B_-^{({\tiny V})}F_{2b}(m_f,m_{\tiny V}), \quad
		f_{ij}^0(m_{\tiny V},m_f)=-i(2-D)A_+^{({\tiny V})}B_+^{({\tiny V})}F_{2b}(m_f,m_{\tiny V}), \nonumber\\
		&&c_{ij}^1(m_{\tiny V},m_f)=-i4\frac{m_f}{m_j}A_+^{({\tiny V})}B_-^{({\tiny V})}F_{3a}(m_f,m_{\tiny V}), \quad
		d_{ij}^1(m_{\tiny V},m_f)=-i4\frac{m_f}{m_i}A_-^{({\tiny V})}B_+^{({\tiny V})}F_{3a}(m_f,m_{\tiny V}), \nonumber\\
		&&e_{ij}^1(m_{\tiny V},m_f)=i2 A_-^{({\tiny V})}B_-^{({\tiny V})}F_{3b}(m_f,m_{\tiny V}), \quad
		f_{ij}^1(m_{\tiny V},m_f)=i2 A_+^{({\tiny V})}B_+^{({\tiny V})}F_{3b}(m_f,m_{\tiny V}).
		\label{coeff_v_opt}
	\end{eqnarray}
	
	The loop integrals $F_{2a}, F_{2b}, F_{3a}$, and $F_{3b}$ are defined as
	\begin{eqnarray}
		&&F_{2a}(m_1,m_2) = (\mu_w)^{2\epsilon}\int_0^1 dx \int \frac{d^D Q}{(2\pi)^D} \frac{1}{(Q^2 + x m_1^2 + (1-x) m_2^2)^2}, \nonumber\\
		&&F_{2b}(m_1,m_2) = (\mu_w)^{2\epsilon}\int_0^1 dx \int \frac{d^D Q}{(2\pi)^D} \frac{1-x}{(Q^2 + x m_1^2 + (1-x) m_2^2)^2}, \nonumber\\
		&&F_{3a}(m_1,m_2) = (\mu_w)^{2\epsilon}\int_0^1 dx \int \frac{d^D Q}{(2\pi)^D} \frac{2x(1-x)}{(Q^2 + x m_1^2 + (1-x) m_2^2)^3}, \nonumber\\
		&&F_{3b}(m_1,m_2) = (\mu_w)^{2\epsilon}\int_0^1 dx \int \frac{d^D Q}{(2\pi)^D} \frac{2x(1-x)^2}{(Q^2 + x m_1^2 + (1-x) m_2^2)^3}.
		\label{loopfunc_opt}
	\end{eqnarray}
	
	Similarly, for scalar boson loops, the amplitude reads
	\begin{eqnarray}
		&&\text{Amp}_{S}(k)=(\mu_{w})^{2\epsilon}\int\frac{d^DQ}{(2\pi)^D}(iA_{\sigma_1}^{(
			{\tiny S})}\omega_{\sigma_1})\frac{i(/\!\!\!Q+/\!\!\!k+m_f)}
		{(Q+k)^2-m_f^2}(iB_{\sigma_2}^{({\tiny S})}\omega_{\sigma_2})
		\frac{i}{Q^2-m_{{\tiny S}}^2}\nonumber \\
		&&\hspace{1.5cm}=i\int_0^1dx\int\frac{d^DQ}{(2\pi)^D}\frac{1}{(Q^2
			+xm_f^2+(1-x)m_{\tiny S}^2)^2}\Big\{1+\frac{2x(1-x)k^2}{Q^2
			+xm_f^2+(1-x)m_{\tiny S}^2}\Big\}\nonumber \\
		&&\hspace{2.0cm}\Big\{A_{\bar{\sigma}}^{({\tiny S})}
		B_{\sigma}^{({\tiny S})}(1-x)/\!\!\!k\omega_{\sigma}+m_f
		A_{\sigma}^{({\tiny S})}B_{\sigma}^{({\tiny
				S})}\omega_{\sigma}\Big\},
		\label{amps1}
	\end{eqnarray}
	where the vertices $A_{\sigma}^{({\tiny S})}$ and $B_{\sigma}^{({\tiny S})}$ are also obtained via SARAH, and $m_{\tiny S}$ denotes the scalar boson mass.  
	The corresponding coefficients $c_{ij}^{0,1}, d_{ij}^{0,1}, e_{ij}^{0,1}, f_{ij}^{0,1}$ follow analogously from Eqs.~(\ref{onel1}) and (\ref{series}), replacing the vector boson masses and vertices with the scalar ones.

	In this work, the mixing of ${\nu_i}^0-{\nu_j}^0$ originates from the following types of one-loop diagrams:
	
	\begin{itemize}
		\item Gauge boson loops involving the $Z$ boson with neutrinos 
		$\nu_{\alpha}^0$ ($\alpha=1,2,\cdots,6$) and the $W$ boson with charged leptons $e_\alpha$ ($\alpha=1,2,3$).
		
		\item Higgs boson loops, including CP-even Higgs bosons $H_\beta^0$ and CP-odd Higgs bosons $A_\beta^0$ 
		($\beta=1,2,\cdots,5$) with neutrinos $\nu_{\alpha}^0$ 
		($\alpha=1,2,\cdots,6$), and charged Higgs bosons $H_\beta^+$ 
		($\beta=1,2,\cdots,4$) with charged leptons $e_\alpha$ 
		($\alpha=1,2,3$).
		
		\item Supersymmetric scalar loops involving CP-even sneutrinos $\tilde{\nu}^R_\beta$ and CP-odd sneutrinos $\tilde{\nu}^I_\beta$ 
		($\beta=1,2,\cdots,6$) with neutralinos $\tilde{\chi}_{\alpha}$ ($\alpha=1,2,\cdots,7$), and sleptons $\tilde{e}_\beta$ ($\beta=1,2,\cdots,6$) with charginos $\chi_\alpha^-$ ($\alpha=1,2,3$).
	\end{itemize}
	
	Then, the one-loop corrections to the neutrino mass matrix elements can be obtained
	\begin{eqnarray}
		\delta m_{\nu,ij}^{1\text{-loop}}
		&=& \delta m_{\nu,ij}^{(Z,\nu^0)}
		+ \delta m_{\nu,ij}^{(W,e^-)}
		+ \delta m_{\nu,ij}^{(H^0,\nu^0)}
		+ \delta m_{\nu,ij}^{(A^0,\nu^0)} \nonumber \\
		&& + \delta m_{\nu,ij}^{(\tilde{\nu}^I,\tilde{\chi})}
		+ \delta m_{\nu,ij}^{(\tilde{\nu}^R,\tilde{\chi})}
		+ \delta m_{\nu,ij}^{(H^+,e^-)}
		+ \delta m_{\nu,ij}^{(\tilde{e},\chi^-)} .
		\label{SUMM}
	\end{eqnarray}
	
	We will briefly outline in the subsequent analysis how our neutrino masses are obtained. First, following Eq.~\ref{meff1} and incorporating experimental neutrino data, we fix the value of $Y_L$. Since $M_R$ is already determined, and the parameters $\mu$ and $\lambda$ are fixed, the value of $v_s$ is thus known. This allows us to determine the expression for $Y_R$. The neutrino data are taken from the PDG~\cite{PDG}, enabling us to solve the relevant equations to obtain the value of $Y_D$, thereby yielding the tree-level neutrino masses. With the tree-level masses in hand, we can further incorporate the one-loop effects via Eq.~\ref{meff1}. To analyze the impact of the one-loop corrections on the neutrino masses, we proceed with the following formula:
	\begin{eqnarray}
		C_{\text{ratio}} = \frac{m_{\nu_{\text{lightest}}} - m_{\nu_{\text{lightest}}}^\text{tree}}{m_{\nu_{\text{lightest}}}} \,,
		\label{CR}
	\end{eqnarray}
	where $m_{\nu_{\text{lightest}}}$ denotes the lightest neutrino mass in the NH or IH mass ordering, and $m_{\nu_{\text{lightest}}}^\text{tree}$ represents its corresponding tree-level value. Equation~(\ref{CR}) thus quantifies the relative effect of one-loop corrections on the lightest neutrino mass. In our analysis, we set $m_{\nu_{\text{lightest}}}^\text{tree} = 0.01~\text{eV}$ for both NH and IH scenarios, while the oscillation parameters $s_{12}^2,\, s_{13}^2,\, s_{23}^2,\, \Delta m_{21}^2,\, |\Delta m_{32}^2|$ are fixed to their central values.
	\begin{figure}[h]
		\setlength{\unitlength}{1mm}
		\centering
		\includegraphics[width=2.5in]{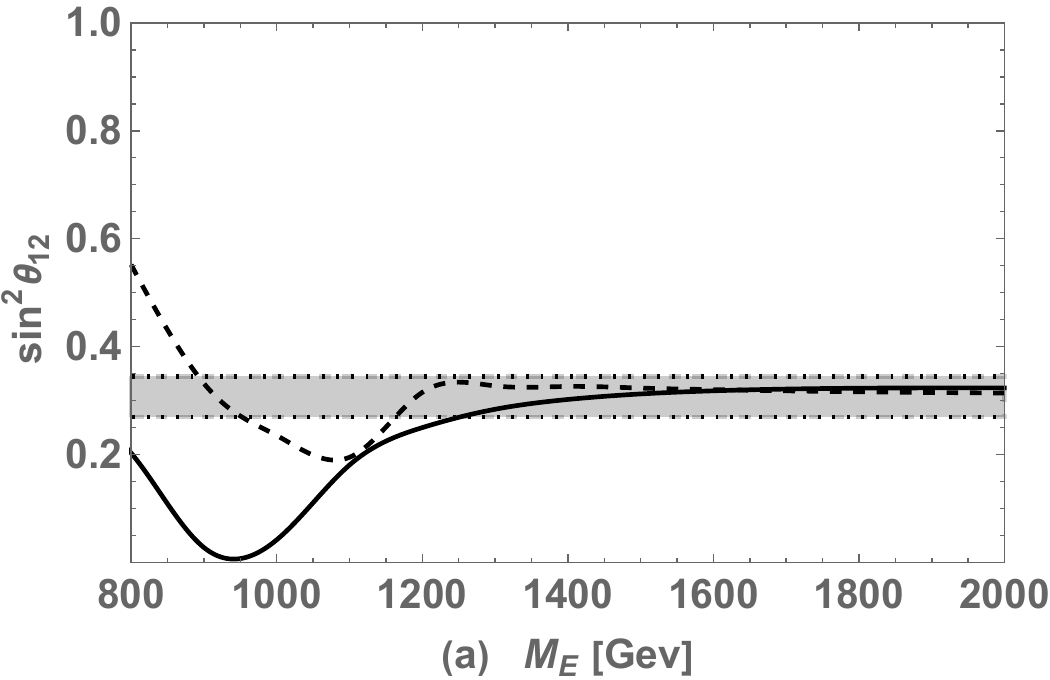}
		\vspace{0.0cm}
		\includegraphics[width=2.5in]{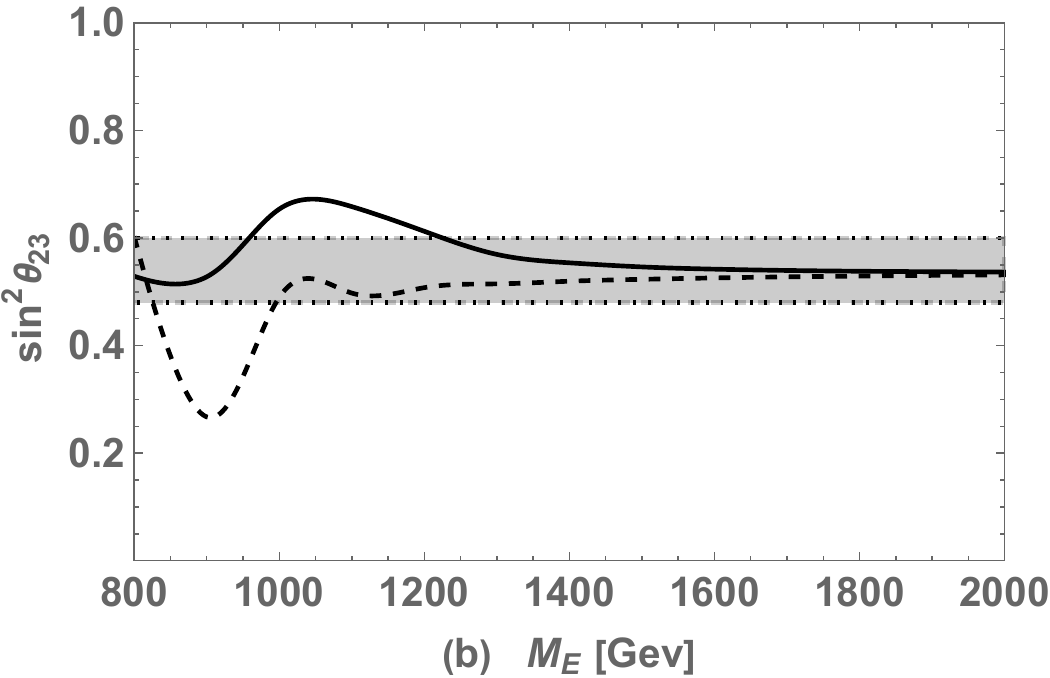}
		\vspace{0.0cm}
		\includegraphics[width=2.5in]{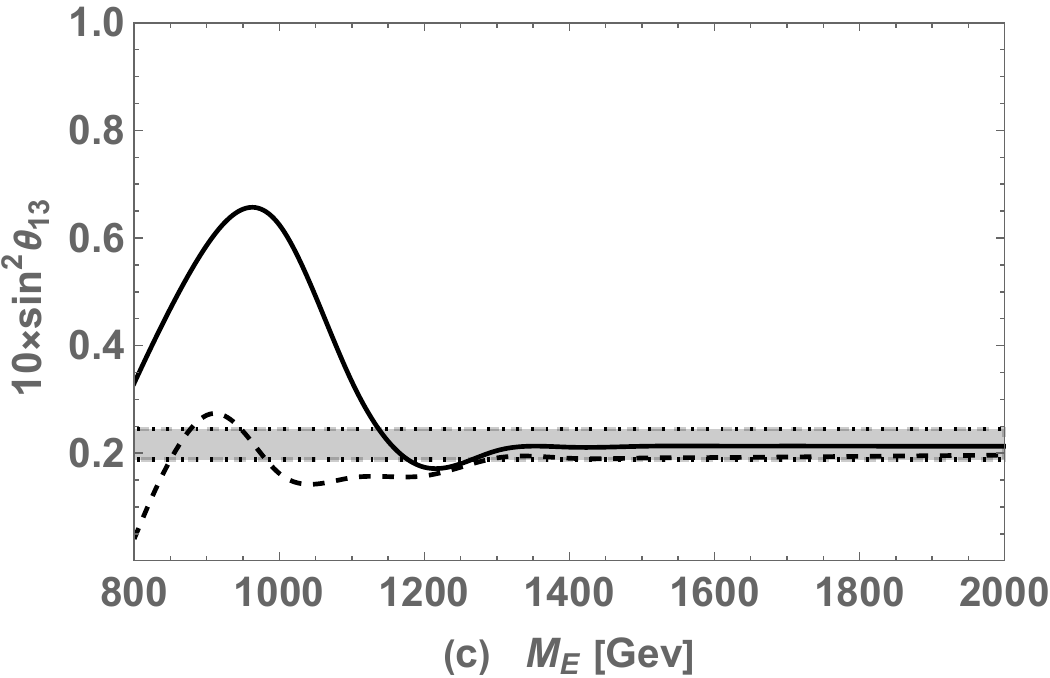}
		\vspace{0.0cm}
		\includegraphics[width=2.5in]{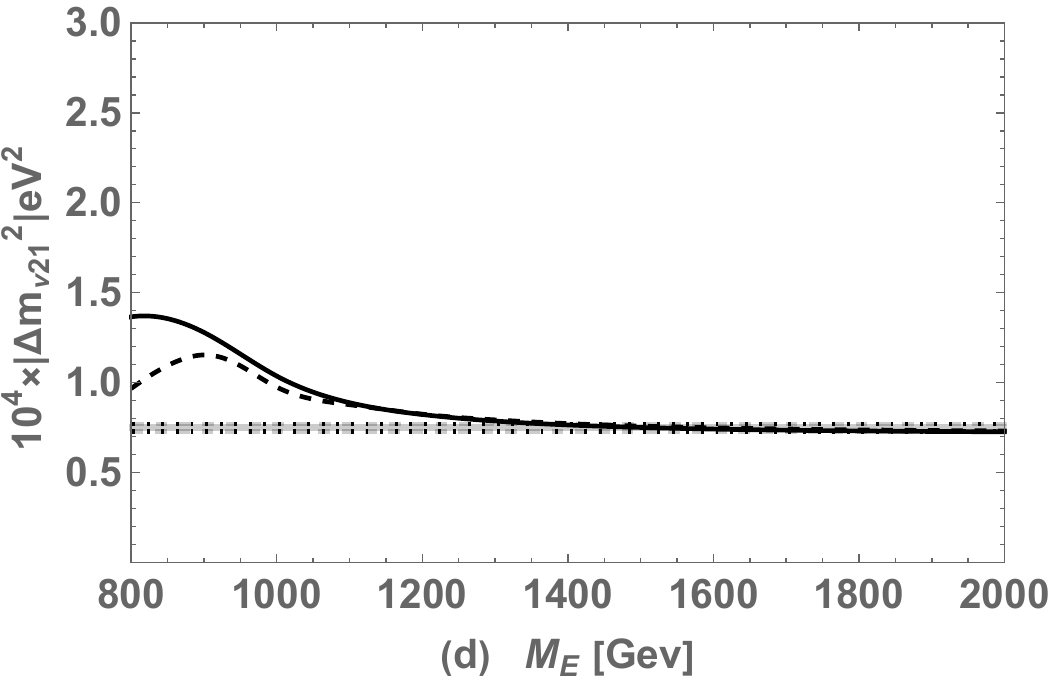}
		\vspace{0cm}
		\vspace{0.0cm}
		\includegraphics[width=2.5in]{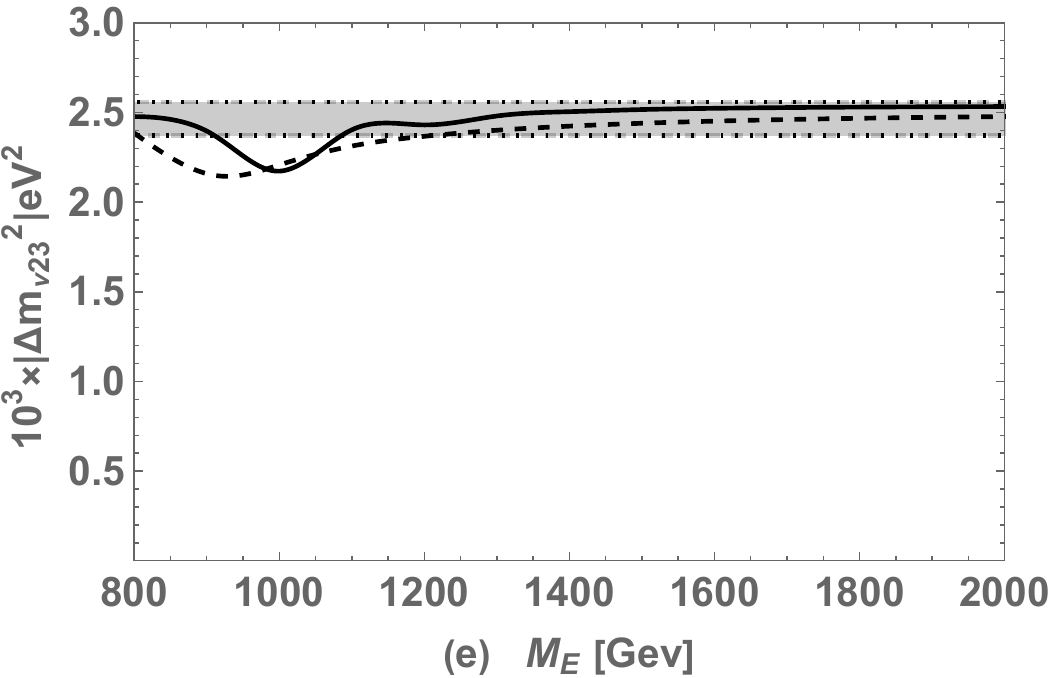}
		\vspace{0.0cm}
		\includegraphics[width=2.5in]{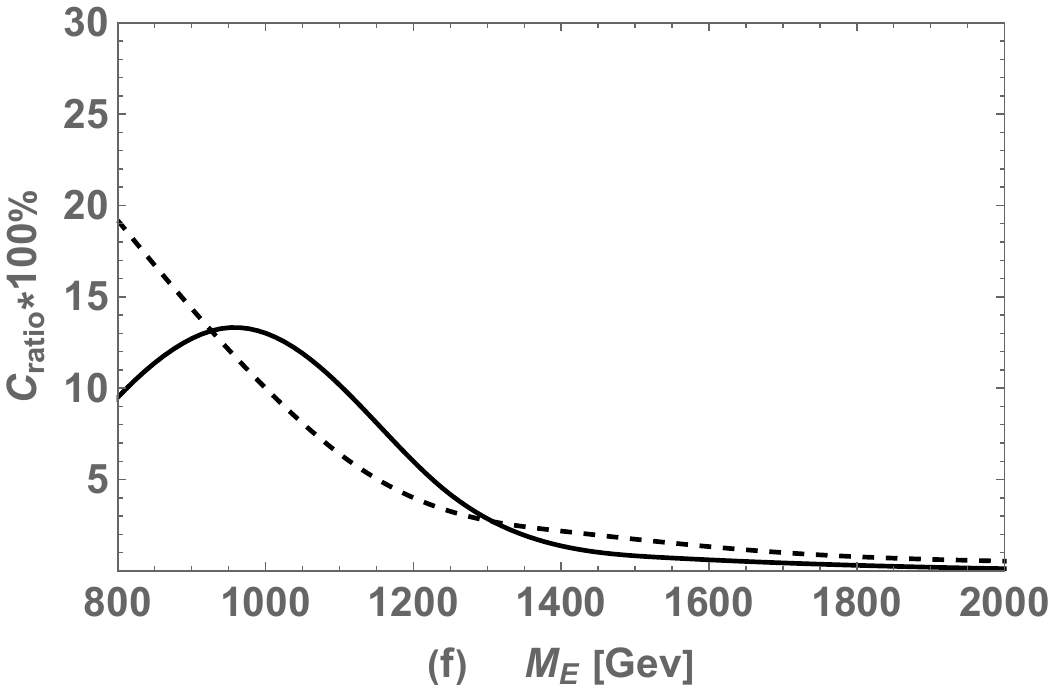}
		\caption{The dependence of the neutrino mixing angles and the neutrino mass-squared differences on the parameter $M_E$ is illustrated in the figures. Panels (a), (b), and (c) correspond to the mixing angles $\sin^2\theta_{12}$, $\sin^2\theta_{13}$, and $\sin^2\theta_{23}$, respectively, while panels (d) and (e) show the mass-squared differences $\Delta m_{\nu_{21}}^2$ and $|\Delta m_{\nu_{32}}^2|$. The gray bands indicate the experimental bounds from neutrino oscillation data. Panel (f) presents the ratio of the one-loop radiative correction as a function of $M_E$, where the solid line corresponds to the IH and the dashed line to the NH.
		}
		\label{TBYLI}
	\end{figure}
	
	Figure~\ref{TBYLI} illustrates the variation of the neutrino mixing angles and mass-squared differences with respect to the parameter $M_E$. In this figure, we take , $\mu = 1000~\text{GeV}$, $M_{2} = 1000~\text{GeV}$. Panels~(a)–(c) show the mixing angles $\sin^2\theta_{12}$, $\sin^2\theta_{13}$, and $\sin^2\theta_{23}$, respectively, while panels~(d) and~(e) present the mass-squared differences $\Delta m_{\nu_{21}}^2$ and $|\Delta m_{\nu_{32}}^2|$. The gray bands in these plots represent the $1\sigma$ experimental bounds obtained from neutrino oscillation data, serving as a reference for the allowed parameter space. Panel~(f) depicts the ratio of the one-loop radiative correction to the lightest neutrino mass, defined in Eq.~(\ref{CR}), as a function of $M_E$. The solid and dashed curves correspond to the inverted IH and the NH, respectively. These results allow us to directly compare the theoretical predictions with current experimental constraints and to assess the impact of varying $M_E$ on both the tree-level and loop-corrected neutrino observables. As $M_E$ increases, both the neutrino mixing angles and the neutrino mass-squared differences exhibit noticeable variations. The three mixing angles are clearly correlated with one another, and a similar correlation is observed between the two mass-squared differences. The parameter $M_E$ enters the mass matrices of sleptons, CP-even sneutrinos, and CP-odd sneutrinos. As $M_E$ increases, the masses of these particles become larger, which in turn suppresses the one-loop contributions. From panel~(f), it can be clearly seen that the ratio defined in Eq.~(\ref{CR}) decreases with increasing $M_E$ for both the IH and NH neutrino mass orderings. A small oscillatory behavior appears in the IH case, which originates from the interference between different loop diagrams. The figure indicates that to ensure the neutrino masses, including one-loop radiative corrections, are consistent with the experimental limits from neutrino oscillation measurements, the parameter $M_E$ must be raised to values exceeding roughly 1300~GeV.
	
	\section{The Wilson coefficients of the process $l_j^-\rightarrow l_i^-\gamma$ and $l_j^-\rightarrow l_i^-l_i^-l_i^+$. \label{wilsonllr}}
	The coefficients corresponding to the Feyaman diagrams shown in Fig.~\ref{FeynmanA}(a) and (b) can be expressed as
	\begin{eqnarray}
		&&A_1^{(a)L} = \frac{1}{6 m_W^2} C_{\bar l_i F^n_k S^c_l}^L C_{\bar F^n_k S^{c*}_l l_j}^R \, I_4(x_{F^n_k},x_{S^c_l}), \quad
		A_2^{(a)L} = \frac{m_{F^n_k}}{m_{l_j} m_W^2} C_{\bar l_i F^n_k S^c_l}^L C_{\bar F^n_k S^{c*}_l l_j}^L \, [I_3 - I_1](x_{F^n_k},x_{S^c_l}), \nonumber\\
		&&A_{1,2}^{(a)R} = A_{1,2}^{(a)L} (L \leftrightarrow R), \nonumber\\
		&&A_1^{(b)L} = \frac{1}{6 m_W^2} C_{\bar l_i F^c_k S^n_l}^R C_{\bar F^c_k S^n_l l_j}^L \, [I_1 - 2 I_2 - I_4](x_{F^c_k}, x_{S^n_l}), \nonumber\\
		&&A_2^{(b)L} = \frac{m_{F^c_k}}{m_{l_j} m_W^2} C_{\bar l_i F^c_k S^n_l}^L C_{\bar F^c_k S^n_l l_j}^L \, [I_1 - I_2 - I_4](x_{F^c_k}, x_{S^n_l}), \nonumber\\
		&&A_{1,2}^{(b)R} = A_{1,2}^{(b)L} (L \leftrightarrow R),
	\end{eqnarray}
	Here, $x_i = m_i^2/m_W^2$, and $C_{abc}^{L,R}$ represent the constant parts of the interaction vertices for particles $a$, $b$, and $c$, which can be obtained using SARAH~\cite{sarah}. The explicit forms of the loop functions $I_{1,2,3,4}$ and $G_{1,2,3,4}$ are given in Refs.~\cite{HB2014,JL2018,ZJ1}.

	The contributions from N-penguin diagrams can be expressed through the following coefficients:
	\begin{eqnarray}
		F^L &=& \frac{1}{2 e^2} C_{\bar l_i F^n_k S^c_l}^R \, C_{S^{c*}_l N S^c_\beta}^R \, C_{\bar F^n_k S^{c*}_\beta l_j} \, G_2(x_{F^n_k}, x_{S^c_\beta}, x_{S^c_l}) \nonumber\\
		&&+ \frac{m_{F^c_k} m_{F^c_\alpha}}{e^2 m_W^2} C_{\bar l_i S^n_\beta F^c_k}^R \, C_{\bar F^c_k N F^c_\alpha}^L \, C_{\bar F^c_\alpha S^n_\beta l_j}^L \, G_1(x_{S^n_\beta}, x_{F^c_k}, x_{F^c_\alpha}) \nonumber\\
		&&- \frac{1}{2 e^2} C_{\bar l_i S^n_\beta F^c_k}^R \, C_{\bar F^c_k N F^c_\alpha}^R \, C_{\bar F^c_\alpha S^n_\beta l_j}^L \, G_2(x_{S^n_\beta}, x_{F^c_k}, x_{F^c_\alpha}), \nonumber\\
		F^R &=& F^L(L \leftrightarrow R).
	\end{eqnarray}
	Here, the loop functions $G_1$ and $G_2$ originate from the evaluation of the three-point integrals with different chiral structures and encode the mass dependence of the internal fermions and scalars.

	Similarly, the coefficients for box-type diagrams are given by
	\begin{eqnarray}
		&&B_1^L=\frac{m_{F^n_k}m_{F^n_\alpha}}{e^2m_W^2}G_3(x_{F^n_k},x_{F^n_\alpha},x_{S^c_\beta},x_{S^c_l})C_{\bar l_i S^c_l F^n_k}^LC_{\bar F^n_k S^{c*}_\beta l_j}^LC_{\bar l_i S^c_l F^n_\alpha}^RC_{\bar F^n_\alpha S^{c*}_\beta l_i}^R\nonumber\\
		&&\qquad\quad+\frac{1}{2e^2m_W^2}G_4(x_{F^n_k},x_{F^n_\alpha},x_{S^c_\beta},x_{S^c_l})[C_{\bar l_i S^c_l F^n_k}^RC_{\bar F^n_k S^{c*}_\beta l_j}^LC_{\bar l_i S^c_\beta F^n_\alpha}^RC_{\bar F^n_\alpha S^{c*}_l l_i}^L\nonumber\\
		&&\qquad\quad+C_{\bar l_i S^c_l F^n_k}^LC_{\bar F^n_k S^{c*}_\beta l_j}^RC_{\bar l_i S^c_l\beta F^n_\alpha}^RC_{\bar F^n_\alpha S^{c*}_l l_i}^L]+\frac{1}{2e^2m_W^2}G_4(x_{F^c_k},x_{F^c_\alpha},x_{S^n_\beta},x_{S^n_l})\nonumber\\
		&&\qquad\quad\times C_{\bar l_i S^n_l F^c_k}^RC_{\bar F^c_k S^{n}_\beta l_j}^LC_{\bar l_i S^n_\beta F^c_\alpha}^RC_{\bar F^c_\alpha S^{n}_l l_i}^L,\nonumber\\
		&&B_2^L=-\frac{m_{F^n_k}m_{F^n_\alpha}}{2e^2m_W^2}G_3(x_{F^n_k},x_{F^n_\alpha},x_{S^c_\beta},x_{S^c_l})C_{\bar l_i S^c_l F^n_k}^RC_{\bar F^n_k S^{c*}_\beta l_j}^RC_{\bar l_i S^c_\beta F^n_\alpha}^LC_{\bar F^n_\alpha S^{c*}_l l_i}^L\nonumber\\
		&&\qquad\quad+\frac{1}{4e^2m_W^2}G_4(x_{F^n_k},x_{F^n_\alpha},x_{S^c_\beta},x_{S^c_l})[C_{\bar l_i S^c_l F^n_k}^RC_{\bar F^n_k S^{c*}_\beta l_j}^LC_{\bar l_i S^c_\beta F^n_\alpha}^LC_{\bar F^n_\alpha S^{c*}_l l_i}^R\nonumber\\
		&&\qquad\quad+C_{\bar l_i S^c_l F^n_k}^RC_{\bar F^n_k S^{c*}_\beta l_j}^LC_{\bar l_i S^c_l F^n_\alpha}^RC_{\bar F^n_\alpha S^{c*}_\beta l_i}^L]+\frac{1}{4e^2m_W^2}G_4(x_{F^c_k},x_{F^c_\alpha},x_{S^n_\beta},x_{S^n_l})\nonumber\\
		&&\qquad\quad\times C_{\bar l_i S^n_l F^c_k}^RC_{\bar F^c_k S^{n}_\beta l_j}^LC_{\bar l_i S^n_\beta F^c_\alpha}^LC_{\bar F^c_\alpha S^{n}_l l_i}^R-\frac{m_{F^c_k}m_{F^c_\alpha}}{2e^2m_W^2}G_3(x_{F^c_k},x_{F^c_\alpha},x_{S^n_\beta},x_{S^n_l})\nonumber\\
		&&\qquad\quad\times C_{\bar l_i S^n_l F^c_k}^RC_{\bar F^c_k S^{n}_\beta l_j}^RC_{\bar l_i S^n_\beta F^c_\alpha}^LC_{\bar F^c_\alpha S^{n}_l l_i}^L,\nonumber\\
		&&B_3^L=\frac{m_{F^n_k}m_{F^n_\alpha}}{e^2m_W^2}G_3(x_{F^n_k},x_{F^n_\alpha},x_{S^c_\beta},x_{S^c_l})[C_{\bar l_i S^c_l F^n_k}^LC_{\bar F^n_k S^{c*}_\beta l_j}^LC_{\bar l_i S^c_\beta F^n_\alpha}^LC_{\bar F^n_\alpha S^{c*}_l l_i}^L\nonumber\\
		&&\qquad\quad-\frac{1}{2}C_{\bar l_i S^c_l F^n_k}^LC_{\bar F^n_k S^{c*}_\beta l_j}^LC_{\bar l_i S^c_l F^n_\alpha}^LC_{\bar F^n_\alpha S^{c*}_\beta l_i}^L]+\frac{m_{F^c_k}m_{F^c_\alpha}}{2e^2m_W^2}G_3(x_{F^c_k},x_{F^c_\alpha},x_{S^n_\beta},x_{S^n_l})\nonumber\\
		&&\qquad\quad\times C_{\bar l_i S^n_l F^c_k}^RC_{\bar F^c_k S^{n}_\beta l_j}^RC_{\bar l_i S^n_\beta F^c_\alpha}^LC_{\bar F^c_\alpha S^{n}_l l_i}^L,\nonumber\\
		&&B_4^L=\frac{m_{F^n_k}m_{F^n_\alpha}}{8e^2m_W^2}G_3(x_{F^n_k},x_{F^n_\alpha},x_{S^c_\beta},x_{S^c_l})C_{\bar l_i S^c_l F^n_k}^LC_{\bar F^n_k S^{c*}_\beta l_j}^LC_{\bar l_i S^c_l F^n_\alpha}^LC_{\bar F^n_\alpha S^{c*}_\beta l_i}^L,\nonumber\\
		&&B_{1,2,3,4}^R=B_{1,2,3,4}^L({L\leftrightarrow R}).
	\end{eqnarray}

	\section{The SUSY contributions to the MDM of the muon. \label{au}}
	The one-loop contributions to the MDM corresponding to Fig.~\ref{FeynmanA} can be expressed as
	\begin{eqnarray}
		a_\mu^{(a)} &=& \Re\Big\{ 4 x_\mu [-I_3(x_{F^n_k},x_{S^c_l}) + I_4(x_{F^n_k},x_{S^c_l})] 
		\Big[(C_{\bar\mu S^c_l F^n_k}^L C_{\bar F^n_k S^{c*}_l \mu}^R) + (C_{\bar\mu S^c_l F^n_k}^R C_{\bar F^n_k S^{c*}_l \mu}^L)^*\Big] \nonumber\\
		&&\quad + \sqrt{x_\mu x_{F^n_k}} [-I_3(x_{F^n_k},x_{S^c_l}) + I_4(x_{F^n_k},x_{S^c_l})] \, C_{\bar\mu S^c_l F^n_k}^R C_{\bar F^n_k S^{c*}_l \mu}^R \Big\}, \nonumber\\
		a_\mu^{(b)} &=& \Re\Big\{ x_\mu [-I_1(x_{F^c_k},x_{S^n_l}) + 2 I_3(x_{F^c_k},x_{S^n_l})] 
		\Big[(C_{\bar\mu S^n_l F^c_k}^R C_{\bar F^c_k S^n_l \mu}^L) + (C_{\bar\mu S^n_l F^c_k}^L C_{\bar F^c_k S^n_l \mu}^R)^*\Big] \nonumber\\
		&&\quad + \sqrt{x_\mu x_{F^c_k}} [2 I_1(x_{F^c_k},x_{S^n_l}) - 2 I_2(x_{F^c_k},x_{S^n_l})] \, C_{\bar\mu S^n_l F^c_k}^R C_{\bar F^c_k S^n_l \mu}^R \Big\}.
	\end{eqnarray}
	
	Under the assumptions $m_F = m_{\chi^+_i} = m_{\chi^0_j} \gg m_W$ and $m_F = m_{\chi^+_i} \gg m_h$, the Barr--Zee type contributions to $a_\mu$, shown in the upper row of Fig.~\ref{BZT}, can be approximated as~\cite{ZJY,JL2018}:
	\begin{eqnarray}
		a_\mu^{WW} &=& \frac{G_F m_\mu^2}{192 \sqrt{2} \pi^4} \Big\{ 5 \left(|C_{\bar\chi_j^0 W^- \chi_i^+}^L|^2 + |C_{\bar\chi_j^0 W^- \chi_i^+}^R|^2\right) \nonumber\\
		&&\quad - 6 \left(|C_{\bar\chi_j^0 W^- \chi_i^+}^L|^2 - |C_{\bar\chi_j^0 W^- \chi_i^+}^R|^2\right) 
		+ 11 \, \Re(C_{\bar\chi_j^0 W^- \chi_i^+}^L C_{\bar\chi_j^0 W^- \chi_i^+}^{R*}) \Big\}, \nonumber\\
		a_\mu^{WH} &=& \frac{G_F m_\mu m_W^2 C_{\bar\mu H^- \nu_\ell}^L}{128 \pi^4 m_F g_2} 
		\Big\{ \Big[\frac{179}{36} + \frac{10}{3} J(m_F^2, m_W^2, m_{H^\pm}^2) \Big] 
		\Re(C_{\bar\chi_j^0 W^- \chi_i^+}^L C_{\bar\chi_j^0 W^- \chi_i^+}^L + C_{\bar\chi_j^0 W^- \chi_i^+}^R C_{\bar\chi_j^0 W^- \chi_i^+}^R) \nonumber\\
		&&\quad + \Big[-\frac{1}{9} - \frac{2}{3} J(m_F^2, m_W^2, m_{H^\pm}^2)\Big] 
		\Re(C_{\bar\chi_j^0 W^- \chi_i^+}^L C_{\bar\chi_j^0 W^- \chi_i^+}^R + C_{\bar\chi_j^0 W^- \chi_i^+}^R C_{\bar\chi_j^0 W^- \chi_i^+}^L) \nonumber\\
		&&\quad + \Big[-\frac{16}{9} - \frac{8}{3} J(m_F^2, m_W^2, m_{H^\pm}^2)\Big] 
		\Re(C_{\bar\chi_j^0 W^- \chi_i^+}^L C_{\bar\chi_j^0 W^- \chi_i^+}^L - C_{\bar\chi_j^0 W^- \chi_i^+}^R C_{\bar\chi_j^0 W^- \chi_i^+}^R) \nonumber\\
		&&\quad + \Big[-\frac{2}{9} - \frac{4}{3} J(m_F^2, m_W^2, m_{H^\pm}^2)\Big] 
		\Re(C_{\bar\chi_j^0 W^- \chi_i^+}^L C_{\bar\chi_j^0 W^- \chi_i^+}^R - C_{\bar\chi_j^0 W^- \chi_i^+}^R C_{\bar\chi_j^0 W^- \chi_i^+}^L) \Big\}, \nonumber\\
		a_\mu^{\gamma h} &=& -\frac{G_F m_\mu m_W^2 C_{\bar\mu h^0 \mu}}{32 \pi^4 m_F} 
		\Re(C_{\bar\chi^+_i h^0 \chi^+_j}^L) \Big[ 1 + \ln \frac{m_F^2}{m_h^2} \Big],
	\end{eqnarray}
	where
	\begin{eqnarray}
		J(x,y,z) = \ln x - \frac{y \ln y - z \ln z}{y - z}.
	\end{eqnarray}
	
	The diagrams in the lower row of Fig.~\ref{BZT} yield analogous contributions, with the same functional structure but with internal neutralinos and charginos replaced by singly charged charginos and the model-specific doubly charged fermions $\tilde{\chi}^{++}$.

	\section{The mass of Higgs  \label{higgs}}
	In the basis $(\phi_d,\phi_u,\phi_s,\phi_T,\phi_{\bar{T}})$, the definition of mass squared matrix for neutral Higgs is given by
	\begin{eqnarray}
		m_h^2=\begin{pmatrix}
			m_{\phi_d \phi_d}&m_{\phi_u \phi_d}&m_{\phi_s \phi_d}&m_{\phi_T \phi_d}&m_{\phi_{\bar{T}} \phi_d}\\
			m_{\phi_d \phi_u}&m_{\phi_u \phi_u}&m_{\phi_s \phi_u}&m_{\phi_T \phi_u}&m_{\phi_{\bar{T}} \phi_u}\\
			m_{\phi_d \phi_s}&m_{\phi_u \phi_s}&m_{\phi_s \phi_s}&m_{\phi_T \phi_s}&m_{\phi_{\bar{T}} \phi_s}\\
			m_{\phi_d \phi_T}&m_{\phi_u \phi_T}&m_{\phi_s \phi_T}&m_{\phi_T \phi_T}&m_{\phi_{\bar{T}} \phi_T}\\
			m_{\phi_d \phi_{\bar{T}}}&m_{\phi_u \phi_{\bar{T}}}&m_{\phi_s \phi_{\bar{T}}}&m_{\phi_T \phi_{\bar{T}}}&m_{\phi_{\bar{T}} \phi_{\bar{T}}}\\
		\end{pmatrix}
	\end{eqnarray}
	where
	\begin{align}
		m_{\phi_d \phi_d}&=m_{H_d}^2+\frac{1}{8}(g_1^2+g_2^2)(2v_{\bar{T}}^2-2v_T^2+3v_d^2-v_u^2)\nonumber\\
		&+\sqrt{2}v_T\Re(A_{\chi_d})+\frac{|\lambda|^2}{2}(v_s^2+v_u^2)-v_s v_{\bar{T}} \Re(\chi_d \lambda_T^*)+(3v_d^2+2v_T^2)|\chi_d|^2,\\	
		m_{\phi_d \phi_u}&=-\frac{1}{4}(g_1^2+g_2^2)v_d v_u-\frac{1}{\sqrt{2}}v_s \Re(A_\lambda)
		+\frac{1}{2}v_T v_{\bar{T}} \Re(\lambda\lambda_T^*)+v_d v_u |\lambda|^2\nonumber\\
		&-v_s v_{\bar{T}}\Re(\chi_u \lambda^*)-v_s v_T\Re(\chi_d \lambda^*)-\frac{1}{2}v_s^2\Re(\kappa\lambda^*),\\
		m_{\phi_u \phi_u}&=m_{H_u}^2-\frac{1}{8}(g_1^2+g_2^2)(2v_{\bar{T}}^2-2v_T^2-3v_u^2+v_d^2)\nonumber\\
		&+\sqrt{2}v_{\bar{T}}\Re(A_{\chi_u})+\frac{1}{2}(v_d^2+v_s^2)|\lambda|^2-v_s v_T \Re(\chi_u \lambda_T^*)+(2v_{\bar{T}}^2+3v_u^2)|\chi_u|^2,\\
		m_{\phi_d \phi_s}&=v_d v_s|\lambda|^2-\frac{1}{\sqrt{2}}v_u \Re(A_\lambda)-v_s v_u \Re(\kappa\lambda^*)
		-v_T v_u \Re(\lambda\chi_d^*)-v_{\bar{T}}v_u \Re(\chi_u \lambda^*)-v_d v_{\bar{T}} \Re(\lambda_T \chi_d^*),\\
		m_{\phi_u \phi_s}&=v_s v_u|\lambda|^2-\frac{1}{\sqrt{2}}v_d\Re(A_\lambda)-v_d v_s\Re(\lambda\kappa^*)
		-v_T v_d\Re(\lambda\chi_d^*)-v_d v_{\bar{T}}\Re(\lambda\chi_u^*)-v_T v_u\Re(\chi_u \lambda_T^*),\\
		m_{\phi_s \phi_s}&=m_S^2+\frac{|\lambda_T|^2}{2}(v_T^2+v_{\bar{T}}^2)-v_T 
		v_{\bar{T}}\Re(\kappa\lambda_T^*)+3v_s^2|\kappa|^2-v_d v_u\Re(\lambda\kappa^*)+\frac{1}{2}(v_d^2+v_u^2)|\lambda|^2,\nonumber\\
		&+\sqrt{2}v_s\Re(A_\kappa),\\
		m_{\phi_d \phi_T}&=-\frac{1}{2}(g_1^2+g_2^2)v_d v_u+\frac{1}{2}v_u v_{\bar{T}}\Re(\lambda\lambda_T^*)-v_u 
		v_s\Re(\lambda\chi_d^*)+4v_d v_T|\chi_d|^2+\sqrt{2}v_d\Re(A_{\chi_d}),\\
		m_{\phi_u \phi_T}&=\frac{1}{2}(g_1^2+g_2^2)v_d v_T-v_u v_s\Re(\lambda_T\chi_u^*)-v_d 
		v_s\Re(\lambda\chi_d^*)+\frac{1}{2}v_d v_{\bar{T}}\Re(\lambda\lambda_T^*),
	\end{align}
	\begin{align}
		m_{\phi_s \phi_T}&=v_s v_T|\lambda|^2-\frac{1}{\sqrt{2}}v_{\bar{T}}\Re(A_\lambda)-
		v_s v_{\bar{T}}\Re(\kappa\lambda_T^*)-v_d v_u \Re(\lambda\chi_d^*)-\frac{1}{2}v_u^2\Re(\chi_u\lambda_T^*),\\
		m_{\phi_T,\phi_T}&=m_T^2-\frac{1}{4}(g_1^2+g_2^2)(2v_{\bar{T}}^2-6v_T^2-v_u^2+v_d^2)+2v_d^2|\chi_d|^2
		+\frac{1}{2}(v_s^2+v_{\bar{T}}^2)|\lambda_T|^2,\\
		m_{\phi_d,\phi_{\bar{T}}}&=\frac{1}{2}(g_1^2+g_2^2)v_d v_{\bar{T}}+\frac{1}{2}v_T v_u \Re(\lambda\lambda_T^*)
		-v_d v_s \Re(\chi_d\lambda_T^*)-v_s v_u\Re(\chi_u\lambda^*),\\
		m_{\phi_u,\phi_{\bar{T}}}&=-\frac{1}{2}(g_1^2+g_2^2)v_{\bar{T}}v_u+\sqrt{2}v_u\Re(A_{\chi_u})
		+\frac{1}{2}v_d v_T\Re(\lambda\lambda_T^*)-v_d v_s \Re(\lambda\chi_u^*)+4v_{\bar{T}}v_u|\chi_u|^2,\\
		m_{\phi_s,\phi_{\bar{T}}}&=v_s v_{\bar{T}}|\lambda_T|^2-\frac{1}{\sqrt{2}}v_T\Re(A_\lambda)
		-v_s v_T\Re(\kappa\lambda_T^*)-\frac{1}{2}v_d^2\Re(\lambda_T \chi_d^*)-v_d v_u\Re(\lambda\chi_u^*),\\
		m_{\phi_T,\phi_{\bar{T}}}&=-(g_1^2+g_2^2)v_T v_{\bar{T}}-\frac{1}{\sqrt{2}}v_s \Re(A_\lambda)
		-\frac{1}{2}v_s^2 \Re(\kappa\lambda_T^*)+\frac{1}{2}v_d v_u \Re(\lambda\lambda_T^*)+v_T v_{\bar{T}}|\lambda_T|^2,\\
		m_{\phi_{\bar{T}},\phi_{\bar{T}}}&=m_{\bar{T}}^2+\frac{1}{4}(g_1^2+g_2^2)(v_d^2-v_u^2-2v_T^2+6v_{\bar{T}})+
		\frac{1}{2}(v_s^2+v_T^2)|\lambda_T|^2+2v_u^2|\chi_u|^2.
	\end{align}
	
	This matrix is diagonalized by $Z^{\rm H}$:
	\begin{equation}
		Z^{\rm H} \, m_h^2 \, Z^{\rm H,\dagger} = \mathrm{diag}\!\left(m_{h_1}^2,\, m_{h_2}^2,\, \ldots\right)
	\end{equation}
	
	with
	\begin{eqnarray}
		\phi_d=\sum_j Z_{j1}^{\rm H} h_j,~~\phi_u=\sum_j Z_{j2}^{\rm H} h_j,~~\phi_s=\sum_j Z_{j3}^{\rm H} h_j,~~\phi_T=\sum_j Z_{j4}^{\rm H} h_j,~~\phi_{\bar{T}}=\sum_j Z_{j5}^{\rm H}.\nonumber
	\end{eqnarray}
	From the tadpole equations, the soft mass parameters 
	$m_{H_d}^2$, $m_{H_u}^2$, $m_S^2$, $m_T^2$, and $m_{\bar{T}}^2$ 
	can be determined from the minimization conditions of the scalar potential.
	
	In the TNMSSM framework, the SM-like Higgs boson mass is
	\begin{eqnarray}
		m_h = \sqrt{\bigl(m_{h_1}^{\rm tree}\bigr)^2 + \Delta m_h^2}, \label{hg2l}
	\end{eqnarray}
	where $m_{h_1}^{\rm tree}$ is the tree-level mass of the lightest CP-even Higgs state and $\Delta m_h^2$ enotes the loop-induced radiative correction.
	The dominant two-loop leading-logarithmic contribution is~\cite{HiggsC1,HiggsC2,HiggsC3,HiggsC4}
	\begin{eqnarray}
		\Delta m_h^2 &=& \frac{3\,m_t^4}{4\pi^2 v^2} 
		\left[ \tilde{t} + \frac{1}{2}\tilde{X}_t 
		+ \frac{1}{16\pi^2} 
		\left( \frac{3 m_t^2}{2 v^2} - 32\pi \alpha_s \right) 
		\left( \tilde{t}^2 + \tilde{t}\,\tilde{X}_t \right) \right], \\
		\tilde{t} &=& \log\!\left( \frac{M_S^2}{m_t^2} \right), 
		\qquad \tilde{X}_t = \frac{2\tilde{A}_t^2}{M_S^2} 
		\left( 1 - \frac{\tilde{A}_t^2}{12 M_S^2} \right).
	\end{eqnarray}
	Here $\alpha_s$ is the running strong coupling constant,  
	$M_S = \sqrt{m_{\tilde t_1} m_{\tilde t_2}}$ is the geometric mean of the stop masses,  
	and $\tilde{A}_t = A_t - \mu\cot\beta$ with $A_t \equiv A_{u,33}$.
	
	\section{Mass matrix for Sleptons \label{slepton}}
	The slepton mass matrix is given by:
	\begin{equation}
		M_{\tilde e}^2 = \begin{pmatrix} 
			m_{\tilde{e}_L \tilde{e}_L^*} & \left( -\dfrac{1}{2} v_s v_u \lambda + v_d v_T \chi_d \right) Y_e^\dagger + \dfrac{1}{\sqrt{2}} v_d A_{he}^\dagger \\[10pt] 
			\dfrac{1}{\sqrt{2}} v_d A_{he} + Y_e \left( -\dfrac{1}{2} v_s v_u \lambda^* + v_d v_T \chi_d^* \right) & m_{\tilde{e}_R \tilde{e}_R^*} 
		\end{pmatrix}
	\end{equation}
	
	The matrix elements are defined as:
	\begin{align}
		m_{\tilde{e}_L \tilde{e}_L^*} &= \tfrac{1}{2} v_d^2 Y_{h_e}^\dagger Y_{h_e}
		+ \tfrac{1}{8} \left( -g_2^2 + g_1^2 \right) \mathbf{1} \left( 2v_T^2 - 2v_T^2 - v_u^2 + v_d^2 \right) 
		+ m_{\tilde L}^2, \\[6pt]
		m_{\tilde{e}_R \tilde{e}_R^*} &= \tfrac{1}{2} v_d^2 Y_{h_e} Y_{h_e}^\dagger 
		+ \tfrac{1}{4} g_1^2 \mathbf{1} \left( -2v_T^2 + 2v_T^2 - v_d^2 + v_u^2 \right) 
		+ m_{\tilde e}^2 .	
	\end{align}
	
	This matrix is diagonalized by \( Z^E \):
	\begin{equation}
		Z^E \, M_{\tilde e}^2 \, Z^{E, \dagger} 
		= \mathrm{diag}\!\left(M_{\tilde e_1}^2,\, M_{\tilde e_2}^2,\, \ldots \right).
	\end{equation}

	with
	\begin{equation}
		\tilde{e}_{L,i} = \sum_j Z_{ji}^{E,*} \tilde{e}_j, \qquad
		\tilde{e}_{R,i} = \sum_j Z_{ji}^{E,*} \tilde{e}_j
	\end{equation}
	Here, $Y_{h_e}$ denotes the Yukawa coupling matrix of the charged leptons. 
	The soft SUSY breaking mass parameters $m_{\tilde L}^2$ and $m_{\tilde e}^2$ correspond to 
	the left-handed slepton doublets and right-handed charged sleptons, respectively, 
	and are taken diagonal in the flavor basis. They do not include the left–right mixing 
	contributions proportional to $A_{he}$. The physical slepton masses $M_{\tilde e_i}$ 
	that enter the muon MDM calculation are obtained after diagonalizing the full 
	mass-squared matrix $M_{\tilde e}^2$ with the unitary matrix $Z^E$.

\end{document}